\documentclass[aps,prb,twocolumn,showpacs,superscriptaddress,floatfix]{revtex4-1}
\usepackage[top=0.6in,right=0.59in,left=0.59in,bottom=0.6in,includefoot,includehead]{geometry}
\usepackage{graphics,graphicx,epstopdf,color,times,bm,bbm,dsfont,fancyhdr,amssymb,amsmath,amsfonts,natbib}
\usepackage{tikz}
\usepackage{subfigure}
\newcommand*\circled[1]{\tikz[baseline=(char.base)]{
            \node[shape=circle,draw,inner sep=0.5pt] (char) {#1};}}
\date{\today}
\usepackage{dcolumn}
\usepackage{latexsym,verbatim}
\usepackage{ulem}
\usepackage{fancybox}
\usepackage{enumitem}
\usepackage[version=3]{mhchem} 
\usepackage{dsfont}
\usepackage{url}
\usepackage[linktoc=page,colorlinks=true,linkcolor=blue,citecolor=blue,urlcolor=blue,breaklinks=true]{hyperref}	
\def\bef{\begin{framed}}
\def\eef{\end{framed}}
\def\be{\begin{equation}}
\def\ee{\end{equation}}
\def\ber{\begin{eqnarray}}
\def\eer{\end{eqnarray}}

\def\Piv{\mbox{\boldmath $\Pi$}}
\def\nn{\nonumber}

\def\rv{{\bm r}}
\def\zv{{\bm {\hat z}}}
\def\pv{{\bm p}}
\def\fv{{\bm f}}

\def\vv{{\bm v}}

\def\yv{{\bm y}}

\def\qv{{\bm q}}
\def\Av{{\bm A}}
\def\Bv{{\bm B}}

\def\vv{{\bm v}}

\begin{document}
\title{Hall viscosity and nonlocal conductivity of ``gapped graphene"}
\author{Mohammad Sherafati}
\affiliation{Department of Physics, Truman State University, Kirksville, Missouri 63501, USA}
\author{Giovanni Vignale}
\email{vignaleg@missouri.edu}
\affiliation{Department of Physics $\&$ Astronomy, University of Missouri, Columbia, Missouri 65211, USA}
\begin{abstract}
We calculate the Hall viscosity and the nonlocal (i.e., dependent on wave vector $\bm q$) Hall conductivity of  ``gapped graphene" (a non-topological insulator with two valleys) in the presence of a strong  perpendicular magnetic field.  Using the linear-response theory at zero temperature within the Dirac approximation for the Landau levels, we present analytical expressions for both valley and total Hall viscosity and conductivity up to $\bm q^2$ at all frequencies. Although the final formulas for total Hall viscosity and conductivity are similar to the ones previously obtained for gapless graphene, the derivation reveals a significant difference between the two systems. First of all, both the Hall viscosity and the Hall conductivity vanish when the Fermi level lies in the gap that separates the lowest Landau level in the conduction band from the highest Landau level in the valence band.  It is only when the Fermi level {\it is not} in the gap that the familiar formulas of gapless graphene are recovered.  Second, in the case of gapped graphene, it is not possible (at least within our present approach) to define a single-valley Hall viscosity: this quantity diverges with a strength proportional to the magnitude of the gap.  It is only when both valleys are included that the diverging terms, having opposite signs in the two valleys, cancel out and the familiar result is recovered.  In contrast to this, the nonlocal Hall conductivity is finite in each valley. These results indicate that the Hoyos-Son formula connecting the Hall viscosity to the coefficient of $q^2$ in the small-$q$ expansion of the $q$-dependent Hall conductivity cannot be applied to each valley, but only to the system as a whole. The problem of defining a ``valley Hall viscosity" remains open.  
\end{abstract}

\maketitle

\section{Introduction}

Graphene-based materials have attracted much attention because of their intriguing electronic properties \cite{GrapheneRev}. As long as both time-reversal and inversion symmetries are respected, graphene remains a gapless system with massless quasiparticles. Addition of a ``mass" term to the Hamiltonian of pristine graphene creates a gap in the energy spectrum. There are 36 possible gap-opening instabilities in graphene associated with spin, valley, and superconducting channels \cite{Chamon2012}. In particular, the addition of the so-called ``Semenoff mass" \cite{Semenoff1984} $\Delta\hat \sigma_z$, where $\Delta$ is the strength of the staggered onsite potential between $A$- and $B-$sublattices and $\hat \sigma_z$ is the $z$-component of the pseudospin operator, has widely been investigated. Semenoff mass opens a band gap of energy $E_{\text{gap}}=2 \Delta$ at the Dirac points while preserving the time-reversal symmetry (in the absence of a magnetic field) and breaking the inversion symmetry of pristine graphene. Such a gapped graphene system with broken inversion symmetry can be realized by growing graphene on various substrates resulting in a wide range of substrate-induced band gaps with values of $\Delta$ ranging from 10~meV to several tens meV. For example, for graphene epitaxilally grown on SiC the gap energy was reported to be $E_{\text{gap}}\approx 260$~meV \cite{GrapheneSiC}, and on hexagonal boron nitride the gap energy 
is observed to be about $E_{\text{gap}}\approx 28$~meV \cite{GraphenehBNEXP1}, 30~meV\cite{GraphenehBNEXP2} or 38~meV \cite{GraphenehBNEXP3}. In addition, one can show that the system is non-topological with vanishing Chern number when both valleys are considered. 

Gapped graphene is host to a unique valley Hall effect even in the absence of a magnetic field \cite{ValleyHallEff}. Similar to the gapless graphene in the presence of a perpendicular external magnetic field \cite{Goerbig2011}, the broken time-reversal symmetry in gapped graphene leads to the formation of Landau levels (LLs) \cite{Koshino2010} whose two-component spinor nature has recently been experimentally confirmed \cite{LLsGapGraph2015}. 
In addition, ultraclean samples of graphene have also been shown to support a hydrodynamic regime for the electron flow~\cite{Hydro},  where the resistance arises from the viscosity when adjacent parts of the fluid move with different velocities. Being the coefficient that controls the transport of momentum in a fluid, viscosity is a fourth-rank tensor that connects the stress tensor with the rate of change of the strain tensor according to the formula
\begin{align} 
P_{ij}=\sum_{kl}\eta_{ij,kl}v_{kl}\,,
\end{align} 
where $i,j,k,l$ are Cartesian indices, $v_{kl}=\frac{1}{2}(\partial_k v_l +\partial_l v_k)$ is the symmetrized gradient of the velocity field $\vv$, and the stress tensor $P_{ij}$ is obtained from the derivative of the Hamiltonian with respect to the metric tensor. The viscosity tensor in $d=2$ dimensions is then given by
\begin{align}
\eta_{ij,kl}=&\zeta \delta_{ij}\delta_{kl}+\eta(\delta_{ik}\delta_{jl}+\delta_{il}\delta_{jk} - \delta_{ij}\delta_{kl})\nn\\ 
&+\frac{1}{2}\eta_H (\epsilon_{ik}\delta_{jl}+\epsilon_{il}\delta_{jk}+\epsilon_{jk}\delta_{il}+\epsilon_{jl}\delta_{ik})
\end{align}
where $\epsilon_{ij}$ is the rank-2 Levi-Civita tensor and $\eta$ and $\zeta$ are the two transport coefficients, the shear and the bulk viscosities, respectively, which are both \textit{dissipative} in rotationally-invariant systems with preserved time-reversal symmetry. The coefficient in the third term denoted by $\eta_H = \eta_{xx,xy}$, produces a force density $f_i=-\sum_j \partial_j P_{ij}$ proportional to the Laplacian of the velocity field, but perpendicular to the latter, viz., 
\begin{align}\label{Force}
\fv = \eta_H \nabla^2 \vv \times \zv,
\end{align}
where $\zv$ is the unit vector perpendicular to the plane of the two-dimensional fluid. Therefore, $\eta_H$ is a \textit{nondissipative} component of the viscosity tensor for any fluid. As a first attempt for an electron fluid, Avron \textit{et al.} obtained an expression for $\eta_H$ for a two-dimensional electron gas (2DEG) in a perpendicular magnetic field \cite{Avron}. They revealed that $\eta_H$ is proportional to the filling factor of the LLs, suggesting a connection with topology. Later, $\eta_H$ was coined as the \textit{Hall viscosity}, denoted by $\eta_{\text{H}}$, and shown \cite{Read2009} to have a topological significance as, for a rotationally-invariant system, it is proportional to average orbital spin known as Wen-Zee shift \cite{WenZee}, a topological feature a quantum Hall system on a curved space possesses, which stems from a ``gravitational coupling" to the intrinsic geometry of the real-space manifold on which the electrons reside. This topological quantity is represented as an excess carrier number density that can be measured. The relationship between the Hall viscosity and the Wen-Zee shift suggests an interesting possibility to experimentally determine the Hall viscosity \cite{Biswas2016}. Interestingly, such relationship has recently been verified for a synthetic photonic quantum Hall system put on the tip of a cone \cite{PhotonicHallVis}. 

Over the past decade, the Hall viscosity has been studied for various quantum Hall systems \cite{HallVisco} {(for a recent calculation of the unquantized Hall viscosity in a two-dimensional electron liquid see Ref.~\onlinecite{Burmistrov2019}).}  It was pointed out by Hoyos and Son \cite{HoyosSon2012} that for Galilean-invariant systems, the Hall viscosity can be extracted from the $q^2$ term in the expansion of the nonlocal Hall conductivity in powers of wave vector $q$, which suggests an experimental access to the Hall viscosity from the electromagnetic response of the electronic quantum Hall system. Due to its two-dimensional structure and the possibility to place it on a curved surface, graphene is an excellent material for the experimental realization of the Hall viscosity for an electronic quantum Hall system.  Recently, Sherafati \textit{et al.} \cite{Sherafati2016} demonstrated that a relation similar to the one derived by Hoyos and Son holds between the Hall viscosity and the Hall conductivity in monolayer graphene, in spite of the absence of the Galilean invariance. They also obtained an analytical expression for the Hall viscosity in a continuum model of graphene using the linear-response formulation of the Hall viscosity suggested by Tokatly and Vignale \cite{TokatlyVignale2007}. 
\begin{figure}[!htb]
\includegraphics[angle=0,width=0.3 \linewidth]{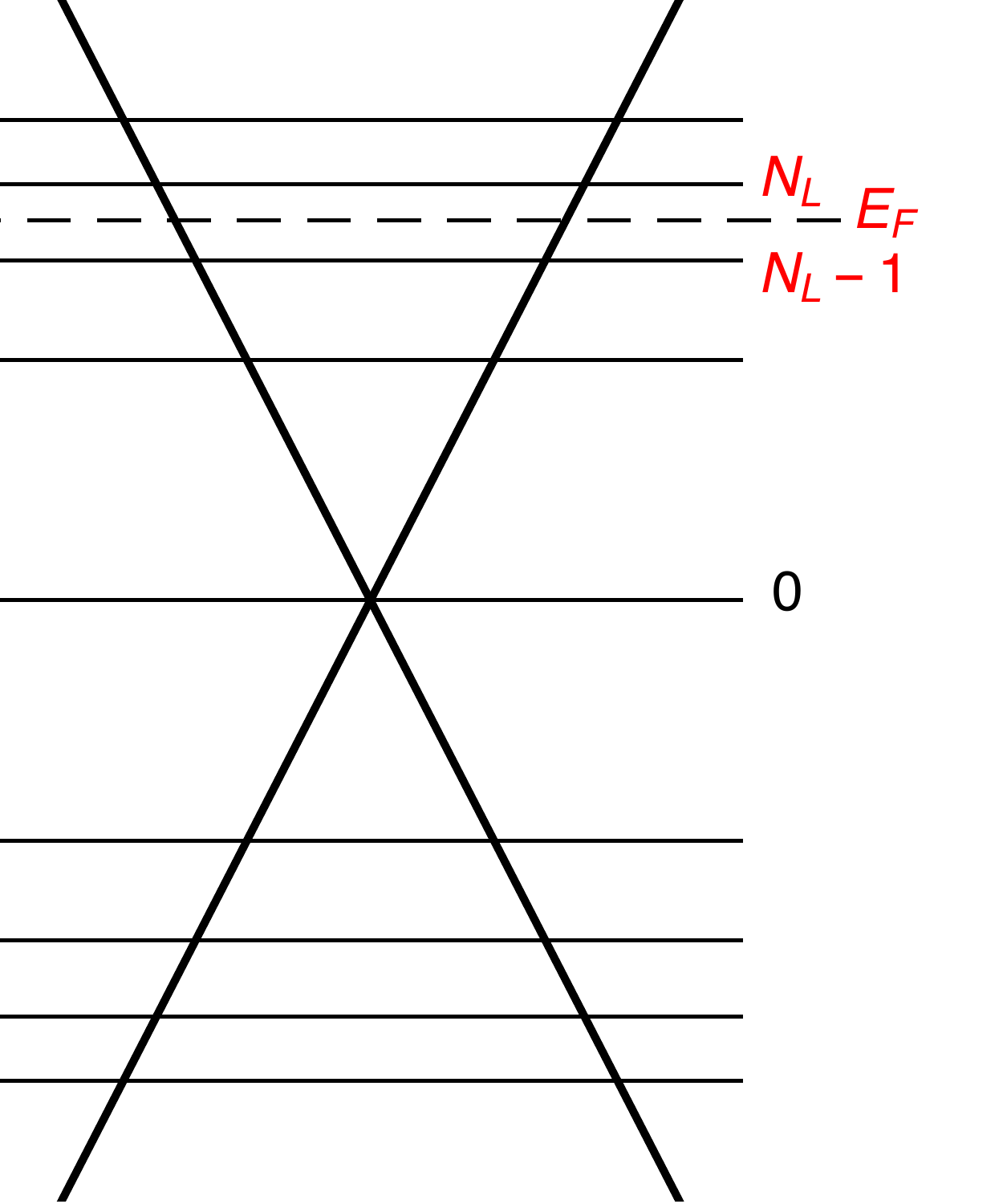}
\caption{Schematic diagram of the Landau levels in gapless graphene. The highest occupied and the lowest empty levels are labeled as $N_L-1$ and $N_L$, respectively, compared to the Fermi level which lies within the gap between these two levels. Here for definiteness we assume electron doping, i.e., $N_L\geq 1$.  However, the final formulas will be given in a form that is invariant under the electron-hole transformation $N_L \to -N_L+1$.} 
\label{Fig1}
\end{figure}

In brief, the quantum Hall viscosity, $\eta_H = \eta_{xx,xy}$, is an instance of a class of ``anomalous transport coefficients" -- of which the Hall conductivity is the best known example -- which are given by the imaginary part of an off-diagonal linear response function, in this case \cite{TokatlyVignale2007}
\be\label{etaHKubo}
 \eta_H =\lim_{\omega \to 0} \Im m \frac{\langle\langle P_{xx};P_{xy}\rangle\rangle_\omega}{\omega}\,,
 \ee 
 where $\langle\langle P_{xx};P_{xy}\rangle\rangle_\omega$ is a shorthand for the off-diagonal stress-stress response function. Starting from the Kubo formula of Eq.~(\ref{etaHKubo}) we obtained \cite{Sherafati2016}
%
\be\label{etaHGraph}
\eta_H = g_{sv}\frac{\hbar}{4\pi \ell^2}\left[N_L^2+(N_L-1)^2\right]{\rm sgn}\left(N_L-\frac{1}{2}\right)\,,
\ee  
($g_{sv}=4$ for graphene) where $\ell=\sqrt{\hbar c/eB}$ is the magnetic length associated with the magnetic field strength $B$ and the Fermi level falls in a gap between the LLs with indices $N_L-1$ and $N_L$ (see Fig.~\ref{Fig1}). $N_L$ can be zero or negative, corresponding to the possibility of hole doping~\cite{GrapheneRev} and the result exhibits full electron-hole antisymmetry, that is to say, the viscosity changes sign under the transformation $N_L \to -N_L+1$.
%
In addition, we found that the expansion of the $q$-dependent Hall conductivity in graphene can be cast in a form very similar to that of found by Hoyos and Son \cite{HoyosSon2012} for 2DEG, namely
\begin{align}\label{GrapheneConductivity1}
\sigma_{xy}(q)\simeq &g_{sv}\left(N_{L}-\frac{1}{2}\right) \frac{e^2}{h}\nn \\
&\times\left\{1 + q^2\ell^2\left[\frac{\vert\eta_H\vert}{4\hbar n} - \left\vert N_{L}-\frac{1}{2}\right\vert\right]\right\}\,,
\end{align}
where $n=g_{sv}\frac{|N_L-1/2|}{2\pi \ell^2}$ is the carrier density (electron density for $N_L\geq 1$, hole density for $N_L \leq0$). Furthermore, we will show that the second term in Eq~(\ref{GrapheneConductivity1}) retains the physical interpretation proposed by Hoyos and Son, i.e., can be expressed in terms of the orbital magnetic susceptibility with the appropriate effective mass. The conclusion is that electrons in doped graphene behave (not unexpectedly) like a Galilean-invariant 2DEG with an effective mass $m_c$ which is given by \begin{align} \label{CyclotronMass}
m_c \equiv \frac{\hbar k_{\text{F}}}{v_{\text{F}}}=\frac{\hbar \sqrt{2|N_L-1/2|}}{v_{\text{F}}\ell}.
\end{align} 
The goal of this work is to extend the calculation of the Hall viscosity and nonlocal Hall conductivity to the case of  ``gapped graphene" (defined in the next section) in a strong magnetic field, and to obtain expressions for the Hall viscosity and Hall conductivity of this system.   Our main results can be summarized as follows:
\begin{enumerate}
\item Both the Hall viscosity and the Hall conductivity vanish when the Fermi level lies in the gap that separates the lowest Landau level in the conduction band from the highest Landau level in the valence band. 

\item When the Fermi level {\it is not} in the gap, the familiar formulas of gapless graphene are recovered.  

\item It is not possible, within our present approach, to define a single-valley Hall viscosity: this quantity diverges with a strength proportional to the magnitude of the gap.  It is only when both valleys are included that the diverging terms, having opposite signs in the two valleys, cancel out and the familiar result is recovered.  

\item In contrast to the above, the nonlocal Hall conductivity is finite in each valley. These results indicate that the Hoyos-Son formula connecting the Hall viscosity to the coefficient of $q^2$ in the small-q expansion of the q-dependent Hall conductivity cannot be applied to each valley, but only to the system as a whole. 

\item The problem of defining a ``valley Hall viscosity" remains open.  
\end{enumerate}

\section{Model and Formulation}
In this work we consider graphene in the presence of a perpendicular magnetic field being gapped due to a Semenoff mass term added to the low-energy Hamiltonian of \textit{both valleys}. The single-particle Hamiltonian for small momenta $\pv$ in $K$- or $K'$-valley is given by
\begin{align}\label{H}
\hat H^\tau=v_{\text{F}}\left(\tau\hat\Piv_x\hat{\sigma}_x+\hat\Piv_y\hat{\sigma}_y\right)+\Delta\hat \sigma_z 
\end{align}
where $\tau=\pm$ is the valley index for $K$ and $K'$ valleys, respectively), $\hat{\sigma}_x$, $\hat \sigma_y$, and $\hat{\sigma}_z$ are pseudospin operators represented by Pauli matrices in the sublattice representation, and $\hat \Piv =\hat \pv+\frac{e}{c}\Av(\hat\rv)$ is the kinetic momentum operator with vector potential $\Av = Bx\hat\yv$ in the Landau gauge corresponding to a perpendicular magnetic field $\Bv=B\hat\zv$.

\begin{figure}[!htb]
\includegraphics[angle=0,width=0.7 \linewidth]{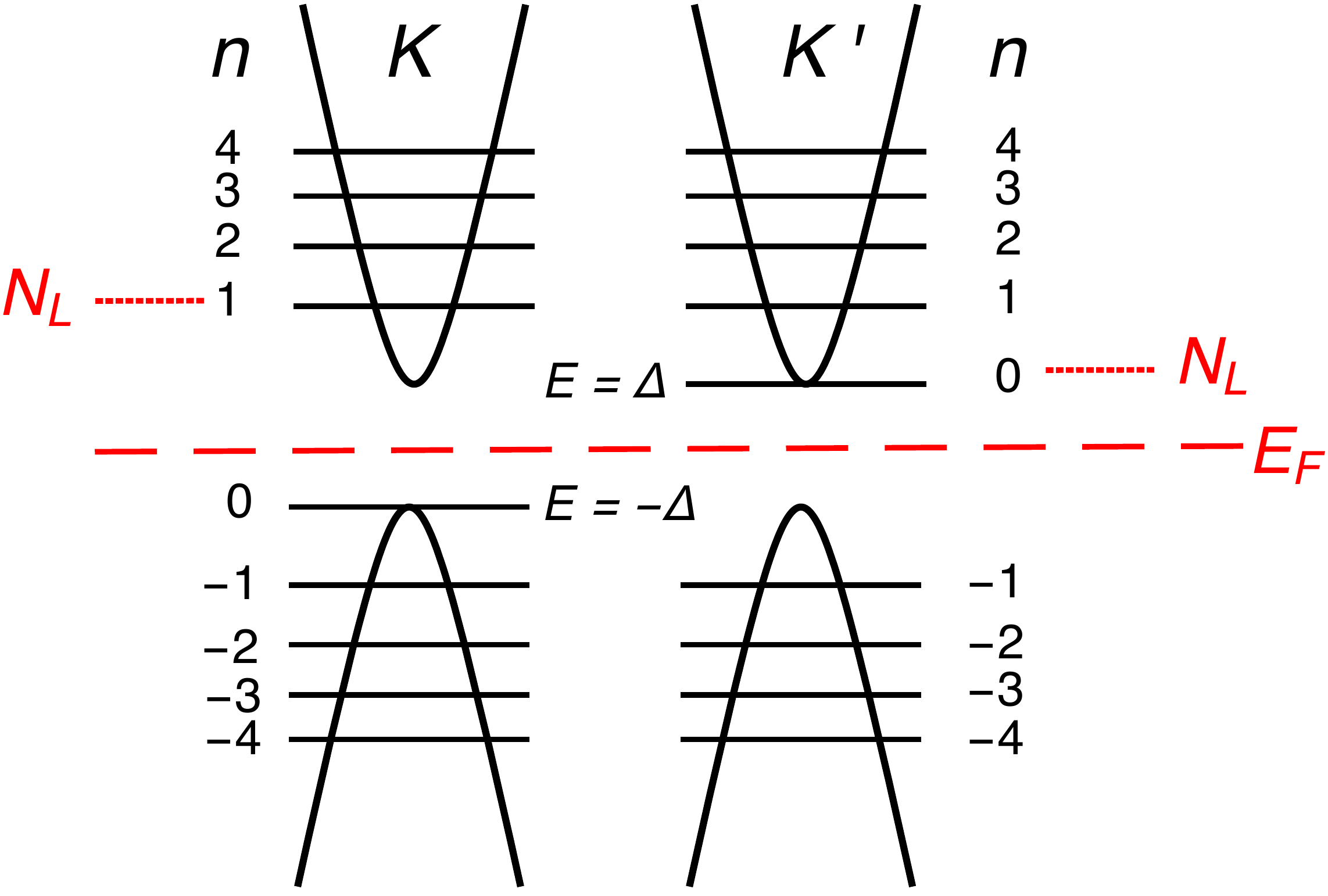}
\caption{Schematic diagram of the Landau levels in ``gapped graphene".  There are two valleys $K$ and $K'$.  The ZLL is at the top of the valence band in the $K$ valley and at the bottom of the conduction band in the $K'$ valley.   The lowest unoccupied Landau level in the $K$ valley is labeled as $N_L$.  This coincides with the lowest unoccupied level in the $K'$ valley  except when the Fermi level is in the gap $|E_{\text{F}}|<\Delta$, in which case $N_L=1$  for $K$-valley and $N_L=0$ for the $K'$-valley.  Here for definiteness we assume electron doping, i.e., $N_L\geq 1$ (for the $K$-valley). However, the final formulas will be given in a form that is invariant under the electron-hole transformation $N_L \to -N_L+1$.} 
\label{fig: LLs}
\end{figure}

The energies of the ``relativistic" Landau levels (LLs) associated with the Hamiltonian in Eq. \eqref{H} for the valence ($\lambda=-$) and conduction ($\lambda=+$) bands are given by


\begin{align}\label{LLs}
E^{\tau}_{\lambda,n}=\begin{cases}
-\tau\Delta&  n=0,\\
\lambda\hbar\omega_0\sqrt{\gamma^2+n} &  n=1, 2, ...,
\end{cases}
\end{align}
where $\omega_0=\sqrt{2}v_{\text{F}}/\ell$ is the characteristic frequency that sets the the Landau energy scale and the distance between the LLs, $v_{\text{F}}$ ($\approx 1\times 10^6$~m/s for gapped graphene on various substrates \cite{LLsGapGraph2015}) is the Fermi velocity. Note the broken valley symmetry manifested by the energy of the Zeroth-Landau Level (ZLL) corresponding to $n=0$. Fig. (\ref{fig: LLs}) shows a schematic diagram of the LLs in gapped graphene for both valleys. The $K$-valley spinor eigenstate is given by   
\begin{align}\label{LLPsi}
\vert\lambda;n,k_y\rangle_{\tau=+}=
\left( \begin{array}{c}\tau\lambda a_{\tau\lambda,n} |n-1,k_y\rangle\\  a_{-\tau\lambda,n}|n,k_y\rangle \end{array}\right)
\end{align}
where and the amplitudes for $\mu=\pm$ are given by 
\begin{align}\label{Amplitude}
a_{\mu,n}=\begin{cases}
\frac{\sqrt{\sqrt{\gamma^2+n}+\mu\gamma}}{\sqrt{2\sqrt{\gamma^2+n}}}&  n\geq 1,\\
0 & \{\mu,n\}=\{-,0\}, \\
1 & \{\mu,n\}=\{+,0\},
\end{cases}
\end{align}
and $\gamma=\frac{\Delta}{\hbar\omega_0}$ is a dimensionless quantity.  The usual Landau-gauge wave functions are given by $\psi_{m,k_y}(x,y)=\langle \rv |m,k_y\rangle =(2^mm!\sqrt{\pi}\ell)^{-1/2} e^{ik_y y} H_m\left(\frac{x}{\ell}+k_y\ell\right) e^{-\frac{(x+k_y\ell^2)^2}{2\ell^2}}$ with $H_m(x)$ being the $m$-th order Hermite polynomial and $|m,k_y\rangle=0$ for $m<0$. In other words, Eqs. \eqref{LLPsi} and \eqref{Amplitude} can also uniquely identify ZLL wavefunction in each valley. The degeneracy per unit area for each Landau level is $(2\pi\ell^2)^{-1}$. 

{We note that $\hat H^\tau$ is connected to  $\hat H^{-\tau}$  by the unitary transformation $\hat U=i\hat \sigma_y$ and a change in sign of $\Delta$
 \be\label{HTransform}
  \hat H^{-\tau}(\Delta)=\hat UH^\tau(-\Delta)\hat U^\dagger\,.
  \ee 
  
Accordingly, the eigenfunction in the $K'$ valley ($\tau=-1$) are obtained from those of the $K$ valley ($\tau=1$) by applying the operator $U=i\sigma_y$ to the latter, and flipping the sign of $\gamma$.  In other words
\be\label{PsiTransform}
\vert\lambda;n,k_y\rangle_{-\tau,\gamma}= \hat U\vert\lambda;n,k_y\rangle_{\tau,-\gamma} 
\ee

The relation between the eigenstates in the $K$ valley and those in the $K'$ valley can be summarized by saying that the conduction band states in valley $K'$ are obtained from the valence band states of valley $K$ by swapping the components of the spinor and flipping the sign of the energy.   Similarly,   the valence band states in valley $K'$ are obtained from the conduction band states of valley $K$ by swapping the components of the spinor and flipping the sign of the energy.}

It is worthwhile justifying the use of the Dirac continuum model. The quantization of the LLs in gapped graphene on SiC has been observed \cite{LLsGapGraph2015} for typical strength of magnetic fields of $B\approx10$~T which corresponds to a magnetic length of $\ell\approx 257 [\text{\AA}]/\sqrt{B[\text{Tesla}]}\approx 81 \ \text{\AA}\gg a$ where $a\simeq 1.42$ \text{\AA} is the carbon-carbon bond length. In addition, at $B=10$~T and for $v_{\text{F}}=10^6$~m/s, we obtain $\hbar \omega_0=114.7$~meV while the Zeeman splitting, $2\mu_{\text{B}}B=1.2$~meV, will be relatively negligible. In other words, each Landau level will be doubly degenerate which allows to introduce a spin multiplicity of $g_{\text{s}}=2$ in the final expressions for the Hall viscosity and conductivity. Finally, referred to studies cited in the Introduction, at $B\approx10$~T, we have $\gamma\simeq 1.1$ for a graphene/SiC ($\Delta=130$~meV) and $\gamma\simeq 0.17$ ($\Delta=19$~meV) for graphene/hexagonal boron nitride substrates, respectively.  

\subsection{Hall Viscosity}

As introduced in the Introduction, the geometric Hall viscosity can be expressed in terms of the matrix elements of the stress tensor \cite{TokatlyVignale2007} as given in Eq. \eqref{etaHKubo}. {The stress tensor operator is defined as \cite{TokatlyVignale2007} 
\begin{align} \label{Stress}
\hat P^\tau_{ij}[g^{ij}]=\frac{2}{\sqrt{g}}\frac{\partial \hat H^\tau}{\partial g^{ij}}
\end{align} 
where $\hat H$ is the many-body Hamiltonian in the presence of a {\it spatially uniform} metric tensor $g^{ij}$~\cite{Sherafati2016}
\ber
\hat H^\tau&=& \frac{v_F}{2}\sum \left\{\sum_{ij} \left(\hat \Pi_i g^{ij}\hat \sigma_j^\tau+\hat \sigma_i^\tau g^{ij}\hat \Pi_j\right)+\Delta\hat\sigma_z\right\}\,.
\eer
The first summation runs over all noninteracting electrons and the second over the indices $i$ and $j$, each taking values $x$ or $y$, with $\hat \sigma_i^\tau=\tau \hat\sigma_x$ if $i=x$,  $\hat \sigma_i^\tau=\hat \sigma_y$ if $i=y$.  $g$ is the determinant of the metric tensor.  
Restricting our attention to a spatially uniform metric tensor  is justified by the fact that we are considering deformations of the electron liquid on a scale much larger than the lattice constant. This restriction dramatically simplifies the Hamiltonian, which would otherwise contain additional spin-connection related to the representation of the Pauli matrices in a curved space time, as discussed in Appendix A of Ref.~\onlinecite{Vozmediano10}.  The spin connection depends on the spatial derivatives of the metric tensor and therefore vanishes in the present case.

Evaluating the derivative of the Hamiltonian with respect to $g^{ij}$  and setting $g^{ij}=\delta_{ij}$ we arrive at the Euclidean stress tensor components 
\begin{align} \label{Pgraphene}
\hat P^{\tau}_{xx}&=2\tau v_{\text{F}}\sum \hat \Pi_x\hat \sigma_x \nn \\
\hat P^{\tau}_{xy}&=v_{\text{F}}\sum \left(\hat \Pi_x\hat \sigma_y+\tau \hat \sigma_x\hat \Pi_y\right),
\end{align} 
Notice that the ``mass term" $\Delta\hat \sigma_z$ does not contribute to the stress tensor.  Because of this fact we have 
\be\label{PTransform}
 \hat P^{-\tau}=\hat U\hat P^\tau\hat U^\dagger\,.
 \ee

Plugging Eq. \eqref{Pgraphene} into Eq. \eqref{etaHKubo} for the matrix elements of the stress tensor between the LLs we obtain the central expression for the Hall viscosity in terms of these matrix elements, which reads

\begin{align}\label{etaH}
\eta^{\tau}_H(\omega)=\frac{2}{\hbar A}\sum_{l,k}^\prime\frac{\Im m \left([\hat P^{\tau}_{xx}]_{kl}[\hat P^{\tau}_{xy}]_{lk}\right)}{\omega^2-(\omega^{\tau}_{lk})^2},
\end{align}
where $A$ is the area of graphene, $\omega^{\tau}_{lk}=\hbar^{-1}(E^{\tau}_{\text{unocc.},l}-E^{\tau}_{\text{occ.},k})$ for the occupied LLs labeled by $k$ and empty levels labeled  by $l$. 

The relation between the Hall viscosity in the $K$ valley and that in the $K'$ valley is directly obtained from the symmetry relations~(\ref{PTransform}) and ~(\ref{PsiTransform}). Changing $\tau \to -\tau$ is equivalent to changing the sign of $\Delta$ in the argument of the sum in Eq.~(\ref{etaH}).  This is because the matrix elements are invariant when the unitary transformation $U$ is applied to both $P_{ij}$ and the eigenstates of the hamiltonian. Also the differences of eigenvalues $\omega^\tau{lk}$ are reversed in sign, but their square is invariant. 
We conclude that 
\be
\eta^{-\tau}_H(\Delta)=\eta^{\tau}_H(-\Delta)\,,
\ee
and therefore the total Hall viscosity can be written as
\be
\eta_H(\Delta)= \eta^{\tau}_H(\Delta)+\eta^{\tau}_H(-\Delta)\,.
\ee
This means that we can limit our calculation to the $K$ valley and simply disregard the terms that are {\it odd} in $\Delta$ (i.e., $\gamma$ in the formulas of our paper).  We will see that the surviving terms are independent of  $\Delta$.  Therefore, the Hall viscosity of gapped graphene generally coincides with the Hall viscosity of gapless graphene. This will be demonstrated by explicit calculation in the next section.}

\subsection{Nonlocal Hall Conductivity}

The valley nonlocal Hall conductivity $\sigma^{\tau}_{xy}(\qv,\omega)$ can be expressed in terms of the current-current response function $\chi^{\tau}_{xy}(\qv,\omega)$ corresponding to that valley, namely 
\begin{align}\label{sigmaXYq}
\sigma^{\tau}_{xy}(\qv,\omega)&=-\frac{1}{\omega}\text{Im}\left[\chi^{\tau}_{xy}(\qv,\omega)\right]. 
\end{align}
The valley response function is given by
\begin{align}\label{Chiqomega}
\chi^{\tau}_{xy}(\qv,\omega)=&\frac{1}{A}\sum_{\substack{{\lambda,\lambda'}\\{n,n'}\\{k_y,k'_y}}}\frac{f(E^{\tau}_{\lambda,n})-f(E^{\tau}_{\lambda',n'})}{\hbar\omega+E^{\tau}_{\lambda,n}-E^{\tau}_{\lambda',n'}+i\hbar0^+} \times \nn \\
&\langle\lambda';n',k'_y\vert\hat{j}^{\tau}_x(\qv)\vert\lambda;n,k_y\rangle \langle\lambda;n,k_y\vert\hat{j}^{\tau}_y(-\qv)\vert\lambda';n',k'_y\rangle
\end{align}
where the wave vector is chosen to be along the $y$-direction ($\qv=q\hat{y}$), $f(E^{\tau}_{\lambda,n})$ is the Fermi distribution function, and the components of the valley current density operator is given by 
\begin{align}\label{Jq}
\hat{j}^{\tau}_x(\qv)=-ev_{\text{F}}\tau\hat{\sigma}_xe^{-i\qv.\hat{\rv}} \nn \\
\hat{j}^{\tau}_y(\qv)=-ev_{\text{F}}\hat{\sigma}_ye^{-i\qv.\hat{\rv}}.
\end{align}

\section{Results}

In this section, we will present the calculations of the Hall viscosity and conductivity. Based on the symmetry, the results for the $K'$-valley can be obtained from those for $K$-valley by replacing $\gamma \to -\gamma$. Therefore, the detailed calculations is only presented for the $K$-valley case.

\subsection{Total Hall Viscosity: Electron-doped Case}

From Eq. \eqref{Pgraphene}, the matrix elements of the single-particle stress tensor for the $K$-valley are given by
\begin{align}\label{PK}
\hat P^K_{xx}=& \frac{\hbar \omega_0}{2}\left(\hat \Pi_+\hat\sigma_+ +\hat \Pi_-\hat\sigma_-\right) +\hat H^K-\Delta \hat\sigma_z \nn \\
\hat P^K_{xy}= & \frac{\hbar \omega_0}{2i}\left(\hat \Pi_+\hat \sigma_+-\hat \Pi_-\hat \sigma_-\right), 
\end{align}
where $\hat \sigma_\pm \equiv \hat \sigma_x\pm i\hat \sigma_y$ and 
\begin{align}
\hat \Pi_\pm= \frac{\ell}{\hbar} \frac{\hat{\Pi}_x\pm i\hat{\Pi}_y}{\sqrt{2}}\,
\end{align}

The expression for the $K$-valley Hall viscosity from Eq. \eqref{etaH} is then simplified to 
\begin{align}\label{etaKH1}
\eta^K_H = -\frac{\hbar \omega_0^2}{4 \pi \ell^2}\sum_{k,l}^\prime\frac{\big|[\Pi_+\sigma_+]_{lk}\big|^2-\big|[\Pi_-\sigma_-]_{lk}\big|^2}{\omega^2-\omega_{lk}^2}
\end{align}

As seen in Fig. \ref{fig: LLs}, for $E_{\text{F}}>-\Delta$, the index of the lowest unoccupied Landau level, $N_L$, starts from one ($N_L\geqslant1$). This index is related to the Fermi energy through the energy spectrum of the LLs, Eq. \eqref{LLs}. This relationship is given by
\begin{align}
N_L=\left\lfloor\left(\frac{E_{\text{F}}}{\hbar\omega_0}\right)^2-\gamma^2\right\rfloor+1,
\label{NL}
\end{align}   
where $\lfloor ...\rfloor$ indicates the floor function. 

The $K$-valley Hall viscosity in Eq. \eqref{etaKH1} is found by taking into account the contributions from three types of transitions: 
\begin{itemize}
\item Case \circled{1}: Both empty and occupied levels belong to the positive sector of the LLs
\item Case \circled{2}: Occupied levels belong to the negative sector and unoccupied levels belong to the positive sector
\item Case \circled{3}: Contribution from the occupied ZLL to the unoccupied levels in the positive sector
\end{itemize}
The first contribution, nonzero only for $N_L\geq 2$, is given by 
\begin{align}
\eta^{K,\circled{1}}_H=-\frac{\hbar \omega_0^2}{\pi \ell^2}\sum_{k=k_{\text{min}}}^{k_{\text{max}}}\frac{a_{+,k+2}^2 a_{-,k}^2(k+1)}{\omega^2-\omega_0^2\left(\sqrt{\gamma^2+k+2}-\sqrt{\gamma^2+k}\right)^2}
\label{etaKCase1}
\end{align}
where $k_{\text{min}}=\text{max}\{1,N_L-2\}$, $k_{\text{max}}=\text{min}\{N_L-1,N_{\text{C}}-2\}$, $N_{\text{C}}$ is the index of the highest-lying Landau level in the positive or negative sector within the linear Dirac bands for which the continuum, Dirac Hamiltonian in Eq. \eqref{H} is valid. This cutoff index can be determined by the condition 

\begin{align}
\hbar\omega_0\sqrt{\gamma^2+N_{\text{C}}}=\hbar v_{\text{F}} k_{\text{C}} \to N_{\text{C}}=\frac{k^2_{\text{C}}\ell^2}{2}-\gamma^2,
\label{Nc}
\end{align} 
where $k_{\text{C}}$ is a cutoff in momentum space. For magnetic fields of the order of $B=10$~T and the ultraviolet momentum cutoff for linear bands in graphene $k_{\text{C}}\sim1/a$ with $a\approx1.42$~\AA, Eq. \eqref{Nc} indicates $N_{\text{C}}\approx1600$ LLs lie within the linear bands. 

It is noteworthy that we insist on keeping $N_{\text{C}}$ as a finite number not infinity in the upper limits of summations for the Hall viscosity, such as in Eq. \eqref{etaKCase1}, and similarly for the nonlocal Hall conductivity such as Eq. \eqref{Ksigmaq4}, in the subsequent sections. This allows us to use properties of summations to simplify the expressions.    

The other two contributions are given by 
\begin{align}\label{etaKCase2}
\eta^{K,\circled{2}}_H=&-\frac{\hbar \omega_0^2}{\pi \ell^2}\sum_{k=k_{\text{min}}}^{N_{\text{C}}-2}\frac{a_{+,k+2}^2 a_{+,k}^2(k+1)}{\omega^2-
\omega_0^2\left(\sqrt{\gamma^2+k+2}+\sqrt{\gamma^2+k}\right)^2}+\nn \\
&\frac{\hbar \omega_0^2}{\pi \ell^2}\sum_{k=N_L}^{N_{\text{C}}-2}\frac{a_{-,k+2}^2 a_{-,k}^2(k+1)}{\omega^2-
\omega_0^2\left(\sqrt{\gamma^2+k+2}+\sqrt{\gamma^2+k}\right)^2},  
\end{align}
and 
\begin{align}\label{etaKCase3}
\eta^{K,\circled{3}}_H (\omega)=&\begin{cases}
-\frac{\hbar \omega_0^2}{\pi \ell^2}\frac{\sqrt{\gamma^2+2}+\gamma}{2\sqrt{\gamma^2+2}}\frac{1}{\omega^2-
\omega_0^2\left(\sqrt{\gamma^2+2}+\gamma\right)^2},& \text{if } N_L=1, 2\\
   0              & \text{otherwise.}
\end{cases}
\end{align}

The sum of the contributions presented in Eqs. \eqref{etaKCase1}, \eqref{etaKCase2}, and \eqref{etaKCase3} yields the final expression of the $K$-valley Hall viscosity. The result for DC Hall viscosity ($\omega=0$) for the $K$-valley is given by
\begin{align}\label{etaK2}
\eta^K_H (\omega=0)=\frac{\hbar}{4\pi \ell^2}\bigg[\sum^{N_L-1}_{k=\text{max}\{0,N_L-2\}}F_1(k,\gamma)+\sum^{N_{\text{C}}-2}_{k=N_L}F_2(k,\gamma)\bigg],
\end{align}
where $N_L\geqslant 1$ and the kernel functions in the summands are given by:

\begin{align}\label{Kfcts}
F_1(k,\gamma)&=\bigg[(k+1)^2-\frac{\gamma(k+1)}{\sqrt{\gamma^2+k+2}}\bigg],\nn \\
F_2(k,\gamma)&=\frac{2\gamma(k+1)}{\sqrt{\gamma^2+k}\sqrt{\gamma^2+k+2}\left(\sqrt{\gamma^2+k+2}+\sqrt{\gamma^2+k}\right)}.
\end{align} 

Similar calculations show that the $K'$-valley Hall viscosity for $N_L\geq 1$ is simply given by the replacement $\gamma \to -\gamma$ in the expression for the $K$-valley in Eq. \eqref{etaK2}. However, for $|E_{\text{F}}|<\Delta$, the ZLL in the $K'$-valley is empty and we must consider all nonzero matrix elements of the stress tensor between this level and occupied LLs in the negative sector. The final expression for the dc $K'$-valley Hall viscosity is given by

\begin{align}\label{etaKp}
\eta^{K'}_H (\omega=0)=\begin{cases}
\eta^K_H (-\gamma,\omega=0) & \text{if } \quad E_{\text{F}}>\Delta \\
\frac{\hbar}{4\pi \ell^2}\sum^{N_{\text{C}}-2}_{k=0}F_2(k,-\gamma)& \text{if} \quad |E_{\text{F}}|<\Delta.
\end{cases}
\end{align}
 
Three remarks are worth making here:
\begin{enumerate}

\item Eq. \eqref{etaK2} is valid for the Fermi level pinned either within the band gap ($N_L=1$) or the gap between any two consecutive LLs in the positive sector ($N_L>1$). 

\item For the infinite number of occupied LLs in the negative sector ($N_{\text{C}} \to\infty$), the second summation in Eq. \eqref{etaK2} diverges. This can be verified by the fact that $F_2(k,\gamma)$ is positive and monotonically decreasing function of $k$ for any $\gamma<1$. Then, the integral test for the convergence of series guarantees the divergence of the second summation. In other words, the single-valley Hall viscosity is diverging for the gapped graphene.  

\item We notice that the gap-dependent terms in Eqs. \eqref{etaK2} and \eqref{etaKp} have opposite signs when replacing $\gamma\to-\gamma$; in particular, $F_2(k,-\gamma)=-F_2(k,\gamma)$. {Therefore, the total dc Hall viscosity, given by the sum of the valley-Hall viscosities,  is finite and independent of the value of the gap, since the diverging gap-dependent terms cancel out in the sum.}
Also, for a Fermi level pinned within the gap corresponding to $N_L=1$ in Eq. \eqref{etaK2}, we use the fact that $F_1(0,\gamma)=F_2(0,\gamma)$ to arrive to the final formula 

\begin{align}\label{etatot}
\eta^{\text{e-doped}}_H(\omega=0)=\begin{cases}
g_{\text{sv}}\frac{\hbar}{4\pi \ell^2}\left[N_L^2+(N_L-1)^2\right]& \text{if } \quad E_{\text{F}}>\Delta \\
0 & \text{if} \quad |E_{\text{F}}|<\Delta.
\end{cases}
\end{align}
where $g_{\text{sv}}=4$ for the valley and spin multiplicity. 
\end{enumerate}

{Eq. \eqref{etatot} shows that the total Hall viscosity is independent of the gap energy. The diverging single-valley contributions cancel out to yield the result for Hall viscosity of the gapless graphene as reported in Eq. (6) of our previous study \cite{Sherafati2016}.  

 The vanishing of the Hall viscosity when the Fermi level, $E_F$, falls in the gap $-\Delta<E_F<\Delta$ follows naturally from these observations.  Indeed, in gapless graphene, the single-valley Hall viscosity is $-1$ (in natural units of $\frac{\hbar}{4\pi \ell^2}$) if $E_F$ is barely below $0$ ($E_F=0^-$)  or $+1$, if it is barely above zero ($E_F=0^+$).  If we now open an infinitesimal gap and set the Fermi level at exactly zero we will be in the case $0^+$ for the $K$ valley and in the case $0^-$ for the $K'$ valley.  Summing the contributions of the two valleys we get $1-1=0$.  Since this result is independent of the magnitude of the gap, we conclude that the Hall viscosity of gapped graphene must be zero when the Fermi level is in the gap.} 
 
The fact that the total Hall viscosity depends only on the position of the Fermi level relative to the Landau levels, and not on the magnitude of the gap, is consistent with the idea that  this quantity is related to a topological index -- the Wen-Zee shift -- as suggested in Ref.~\onlinecite{Golkar2014} for the case of gapless graphene (see, in particular, Eq. 9.2 and the equation for the Wen-Zee shift given after Eq. 5.1).  However, the Hall viscosity evinced from the equations of Ref.~\onlinecite{Golkar2014} would vanish at $n=0$ (corresponding to $N_L=1$ in our paper), which is in contradiction with all the calculations that have been published to date\cite{Sherafati2016,Tuegel2015,Kimura2010} (differences between numerical factors obtained by these authors are discussed in Ref.~\onlinecite{Sherafati2016}).  The reason for the discrepancy is not presently understood.  It may be due to the fact that the connection between the Wen-Zee shift and the Hall viscosity was obtained from a field theoretical approach under the assumption of continuous rotation symmetry (see Ref.~\onlinecite{ReadRezayi2011}, after Eq. 1.8) which is certainly broken in a lattice system such as graphene (and even more so in the presence of disorder).  Even though continuous rotation symmetry emerges in the continuum limit, the generator of the rotations would include a pseudospin term, which makes the connection with field-theoretical results quite a delicate matter.  Calculations of the Wen-Zee shift for lattice system could shed light on this question.

\subsection{Total Hall Viscosity: Hole-doped Case}
The hole doping corresponds to the negative Fermi levels such that ${E_{\text{F}}<-\Delta}$. This corresponds the shift of $N_L$ index for the electron-doped case to $-N_L+1$ for the hole-doped one. In other words, for the latter case, all the LLs in the negative sector below the highest empty one labeled as $-N_L$ are now occupied. As before we will consider three cases:
\begin{itemize}
\item Case \circled{1}: Interband transitions for which both empty and occupied levels belong to the negative sector. In this case we have $-N_L+1\leqslant l \leqslant -1$ for empty levels and $-N_{\text{C}} \leqslant k \leqslant -N_L$ for the occupied ones.   
\item Case \circled{2}: Interband transitions for which the occupied levels belong to the negative sector and the  unoccupied ones belong to the positive sector. In this case we have $1\leqslant l \leqslant N_{\text{C}}$ for empty levels and $-N_{\text{C}} \leqslant k \leqslant -N_L$ for the occupied ones. 
\item Case \circled{3}: Contribution from the occupied LLs in the negative sector to the unoccupied ZLL. In this case we have $l=0$ and $-N_{\text{C}} \leqslant k \leqslant -N_L$ 
\end{itemize}

It turns out that the contribution to the hole-doped $K$-valley Hall viscosity for each of the above cases is exactly the opposite of the corresponding case for the electron-doped $K'$-valley Hall viscosity, and vice versa. In other words, we obtain the following relationships: 
\begin{align}
\eta^{K,\circled{1}}_H (\omega)=-\eta^{K',\circled{1}}_H (\omega)\nn \\
\eta^{K,\circled{2}}_H (\omega)=-\eta^{K',\circled{2}}_H (\omega) \nn \\
\eta^{K,\circled{3}}_H (\omega)=-\eta^{K',\circled{3}}_H (\omega),
\end{align}
where the left-hand side is the contribution for the electron doing and the right-hand sign is that for the hole doping. We conclude that for $N_L\leqslant -1$ the total Hall viscosity changes sign under transformation $N_L \to -N_L+1$ and we must have 
\begin{align}\label{etaHEehSymm}
\eta^{\text{e-doped}}_H(\omega)=-\eta^{\text{h-doped}}_H(\omega).
\end{align}
To incorporate this symmetry, one must multiply the Hall viscosity in Eq. \eqref{etatot} by a factor of ${\rm sgn}\left(N_L-\frac{1}{2}\right)$.
\subsection{Total Nonlocal Hall conductivity: Electron-doped Case}
In the Appendix \ref{AppA}, we have demonstrated that the total nonlocal Hall viscosity vanishes for an electron-doped case where the Fermi level is pinned within the gap, namely for $|E_{\text{F}}|<\Delta$ corresponding to $N_L=1$ for $K$-valley and $N_L=0$ for the $K'$-valley as shown in Fig. \ref{fig: LLs}. In this section, we then discuss the electron-doped case for which $E_{\text{F}}>\Delta$ corresponding to $N_L\geq 1 $ for both valleys. As discussed in Appendix \ref{AppA}, the results for the $K'$-valley are simply found from those for the $K$-valley by the replacement $\gamma \to -\gamma$.  
Using the LL eigenstates from Eq. \eqref{LLPsi} and the $K$-valley current density operator from Eq. \eqref{Jq}, we can finally obtain the matrix elements of $x$- and $y$-components the current density to be plugged in the valley response function in Eq. \eqref{Chiqomega} as
\begin{widetext}
\begin{align} \label{JxJy4K}
\langle\lambda';n',k'_y\vert\hat{j}^K_x(\qv)\vert\lambda;n,k_y\rangle&=-e v_{\text{F}}\bigg[\lambda'a_{-\lambda,n}a_{\lambda',n'}
I(n'-1,n,\qv)+\lambda a_{\lambda,n}a_{-\lambda',n'}I(n',n-1,\qv)\bigg]\nn\\
\langle\lambda;n,k_y\vert\hat{j}^K_y(-\qv)\vert\lambda';n',k'_y\rangle&=-e v_{\text{F}}\bigg[-i\lambda a_{\lambda,n}a_{-\lambda',n'}
I^*(n',n-1,\qv)+i\lambda' a_{-\lambda,n}a_{\lambda',n'}I^*(n'-1,n,\qv)\bigg],
\end{align} 
\end{widetext}
where the integral $I(n',n,\qv)$ is given by
\begin{align}\label{Iq4K}
I(n',n,\qv)&=\int d\rv \psi^*_{n',k'_y}(x,y)e^{-i\qv.\rv}\psi_{n,k_y}(x,y)\nn \\
&=\delta_{q,k_y-k'_y}\times \begin{cases}
G_{n'n}(\qv) & n\geqslant n'\\
 \bigg[G_{nn'}(-\qv)\bigg]^*,              & n<n'
\end{cases}
\end{align}
where the function $G_{n'n}(\qv)$ is given by
\begin{align}\label{Gq4K}
G_{n'n}(\qv)=\sqrt{\frac{n'!}{n!}}\left(-\frac{\ell q}{\sqrt{2}}\right)^{n-n'}e^{-\ell^2 q^2/4}L^{n-n'}_{n'}(q^2\ell^2/2),
\end{align}
where $L^m_n(x)$ is the associated Laguerre polynomial, and for two integers $m<n$, we have also made a use of the integral \cite{GR} $\int_{-\infty}^\infty e^{-x^2}H_m(x+y)H_n(x+z)dx=2^n\pi^{1/2}m!y^{n-m}L^{n-m}_m(-2yz)$. We then plug the matrix elements of Eq. \eqref{JxJy4K} into the expression for the response function [Eq. \eqref{Chiqomega}], use the symmetry properties of $I(n',n,\qv)$ integral, and find a closed summation for the nonlocal Hall conductivity using Eq. \eqref{sigmaXYq}; the details are discussed in Appendix \ref{AppA}. The simplified expression for the nonlocal Hall conductivity is then given by
\begin{align}\label{Ksigmaq2}
\sigma^K_{xy}(\qv,\omega)=\frac{e^2\omega_0^2}{h}\sum_{\substack{{\lambda,n;E_{\lambda,n}<E_{\text{F}}}\\{\lambda',n';E_{\lambda',n'}>E_{\text{F}}}}}\frac{\mathcal{I}^K(\lambda, n;\lambda', n', \qv)}{\omega^2-\omega_{\lambda n,\lambda' n'}^2},
\end{align}   
where $\omega_{\lambda n,\lambda' n'}=(E^K_{\lambda,n}-E^K_{\lambda',n'})/\hbar$ given by Eq. \eqref{LLs} with $\tau=+$ and the function in the numerator is given 
\begin{widetext}
\begin{align} \label{IK}
\mathcal{I}^K(\lambda, n;\lambda', n', \qv)&=\lambda^2 a_{\lambda,n}^2a_{-\lambda',n'}^2\bigg\vert I(n',n-1,\qv)\bigg\vert^2-\lambda'^2 a_{-\lambda,n}^2a_{\lambda',n'}^2\bigg\vert I(n'-1,n,\qv)\bigg\vert^2=-\mathcal{I}^K(\lambda', n';\lambda, n, \qv),
\end{align} 
\end{widetext}
which is antisymmetric with respect to the exchange of $\{\lambda,n\}\longleftrightarrow\{\lambda',n'\}$.

Next, we calculate three contributions to the nonlocal Hall conductivity, similar to the Hall viscosity, and expand the final double summation expressed in Eq. \eqref{Ksigmaq4} up to $q^2$. The details of this calculation is presented in the last section of the appendix. The final result for the AC nonlocal Hall conductivity is given by
\begin{widetext}
\begin{align}\label{ACsigmaK}
\sigma^K_{xy}(q,N_L,\gamma,\Omega)\simeq\sigma_0\bigg[F(N_L,\gamma,\Omega)+(q\ell)^2\bigg(G(N_L,\gamma,\Omega)+G(N_L+1,\gamma,\Omega)-\left(\frac{2N_L-1}{2}\right)F(N_L,\gamma,\Omega)\bigg)\bigg],
\end{align} 
\end{widetext}
where $\sigma_0=\frac{e^2}{2h}$, $\Omega=\omega/\omega_0$, and the functions $F(N_L,\gamma,\Omega)$ and $G(N_L,\gamma,\Omega)$ are given by
\begin{widetext}
\begin{align}\label{ACsigmaKgfcs}
F(N_L,\gamma,\Omega)&\equiv-\left(1+\frac{\gamma}{\sqrt{\gamma^2+N_L}}\right)\frac{\Omega ^2- \left(2\gamma ^2-2\gamma\sqrt{\gamma ^2+N_L}+2N_L-1\right)}{ \Omega ^4-2 \Omega ^2  \left(2 \gamma ^2+2 N_L-1\right)+1}\nn \\
G(N_L,\gamma,\Omega)&\equiv\left(\frac{1-N_L}{2}\right)\left(1+\frac{\gamma}{\sqrt{\gamma^2+N_L}}\right)\frac{\Omega ^2-\left(2\gamma^2-2\gamma\sqrt{\gamma ^2+N_L}+2N_L-2\right)}{\Omega^4-4 \Omega ^2 \left(\gamma ^2+N_L-1\right)+4}.
\end{align}
\end{widetext}
\begin{figure}[htb!]
%
        \subfigure[]{%
            \label{Fig3a}
            \includegraphics[width=0.4\textwidth]{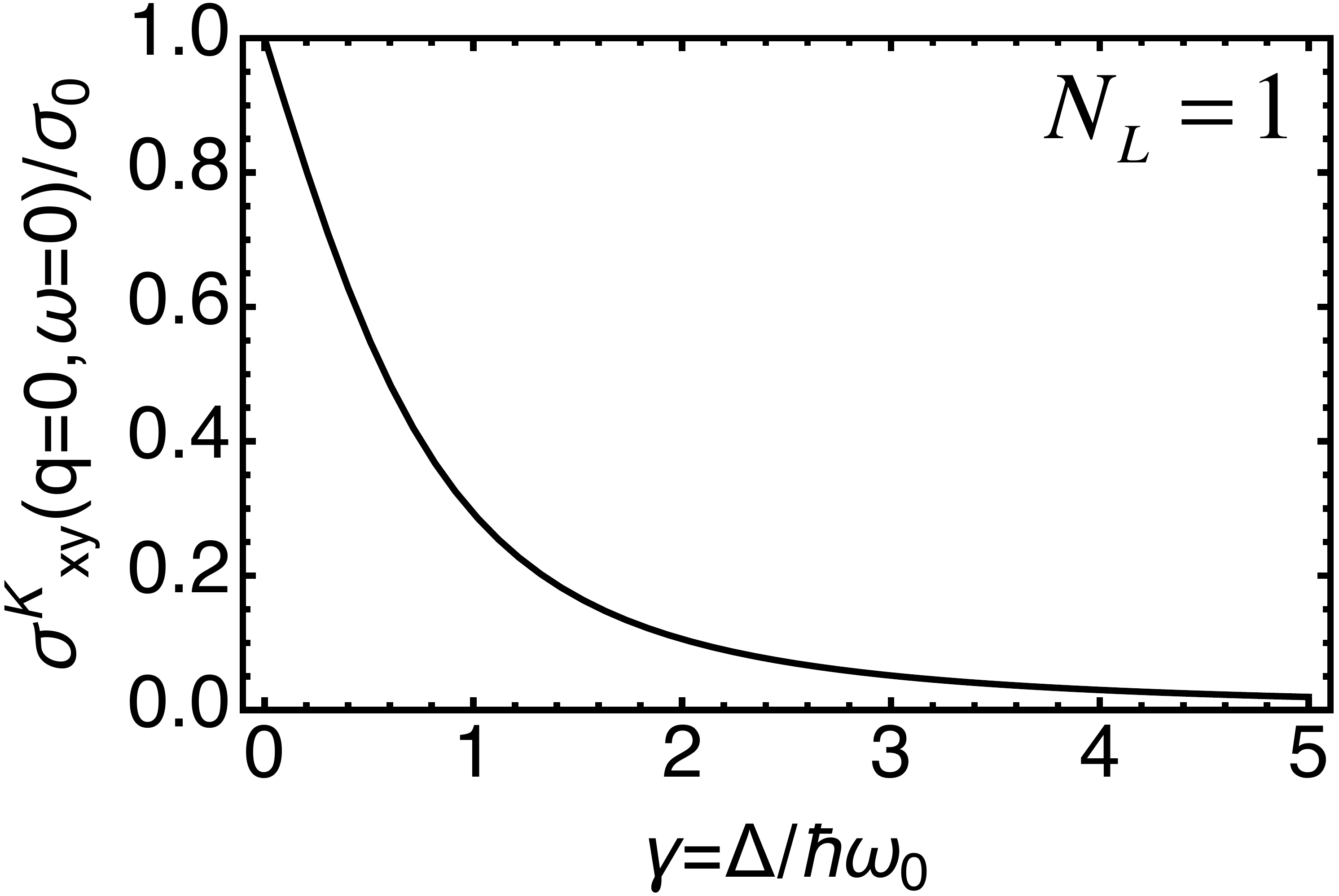}
        }%
        \subfigure[]{%
           \label{Fig3b}
           \includegraphics[width=0.4\textwidth]{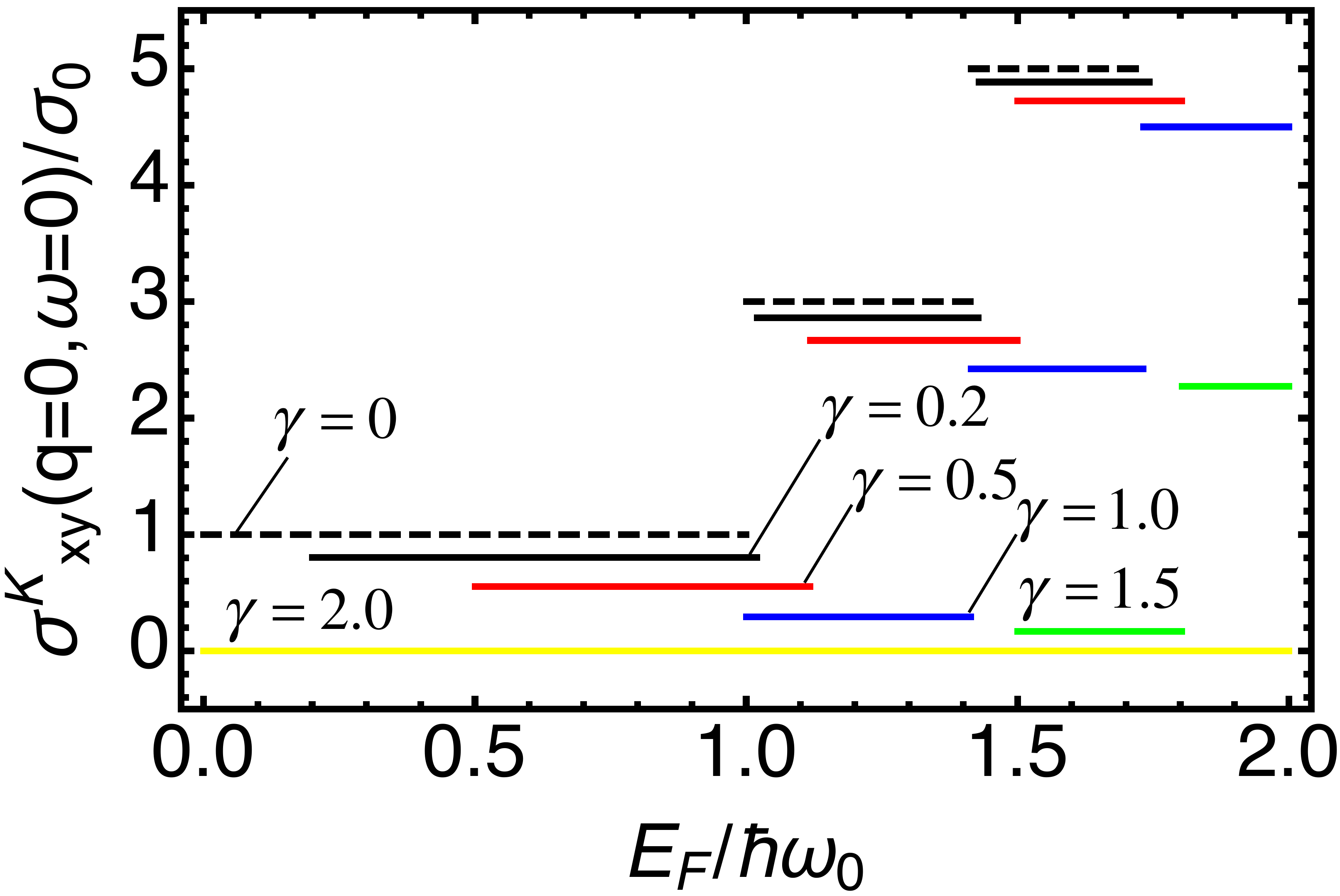}
        } 
       
    \caption{$K$-valley DC Hall conductivity for $q=0$ [Eq. \eqref{DCsigmaK}] as a function of a) energy gap $\gamma$ for $N_L=1$ and b) the Fermi energy evaluated using Eq. \eqref{NL} for several gap values, namely $\gamma=0 , 0.2, 0.5 , 1.0 , 1.5 , 2.0$ indicated by dashed black, solid black, red, blue, green, and yellow lines, respectively. For a magnetic field of $B=10$~T and a Fermi speed of $v_{\text{F}}=10^6$~m/s, these $\gamma$ values correspond to the energy gap values o $E_{\text{gap}}=45.9, 114.7, 229.5, 344.2, 458.9$~meV, respectively.}
\label{Fig3}
\end{figure}
Eq. \eqref{ACsigmaK} along with \eqref{ACsigmaKgfcs} is one of the central results of this paper. Setting $\Omega=0$, we arrive at the DC nonlocal $K$-valley Hall conductivity for the gapped graphene given by 
\begin{widetext}
\begin{align}\label{DCsigmaK}
\sigma^K_{xy}(q,N_L,\gamma,\Omega=0)&\simeq \sigma^{(0),K}_{xy}(q=0,N_L,\gamma,\Omega=0)+\sigma^{(2),K}_{xy}(q,N_L,\gamma,\Omega=0),\nn \\
\text{where}\nn \\
\sigma^{(0),K}_{xy}(q=0,N_L,\gamma,\Omega=0)&=\sigma_0\left(2N_L-1-\frac{\gamma}{\sqrt{\gamma^2+N_L}}\right)\nn \\
\sigma^{(2),K}_{xy}(q,N_L,\gamma,\Omega=0)&=\left(-\frac{\sigma_0q^2\ell^2}{4}\right)\left(1+\frac{\gamma}{\sqrt{\gamma^2+N_L}}(1-3N_L)+\frac{\gamma N_L}{\sqrt{\gamma^2+N_L+1}}-6N_L+6N^2_L\right).
\end{align}
\end{widetext}

Eq. \eqref{DCsigmaK} clearly indicates that the $K$-valley conductivity is no longer quantized for the gapped graphene as it is the case for the gapless system. The continuous decay of the quantity with the gap energy is shown in Fig. \ref{Fig3a}. 
{Setting $\gamma=0$ in Eq. \eqref{DCsigmaK}, we immediately recover the DC $K$-valley Hall conductivity of the gapless graphene given by
\begin{widetext}
\begin{align}\label{DCsigmaKgapless}
\sigma^K_{xy}(q,N_L,\Omega=0)\simeq g_{\text{s}}\frac{e^2}{h}\bigg[N_L-\frac{1}{2}-\frac{q^2\ell^2}{8}\left(6N^2_L-6N_L+1\right)\bigg].
\end{align} 
\end{widetext}  
Eq. \eqref{DCsigmaKgapless} was also obtained in our previous study of the Hall conductivity for the gapless case \cite{Sherafati2016}. Setting $q=0$ in Eq. \eqref{DCsigmaKgapless} recovers the well-known ``anomalous" quantized Hall conductivity of electrons in gapless graphene $\sigma^K_{xy}(q=0,N_L,\gamma=0,\Omega=0)=\sigma_0(2N_L-1)$ which yields odd-integer plateaus in the gapless graphene \cite{AQHGraph} as shown in Fig. \ref{Fig3b} with dashed plateaus. }

Notably, as the gap opens, the conductivity is no longer quantized as in the gapless graphene and the values of the plateaus decrease with increasing gap. The later point is further illustrated in Fig. \ref{Fig4}.

As shown in Appendix \ref{AppA}, the result for the $K'$-valley corresponding to $N_L\geq 1$ can be found from that for the $K$-valley evaluated at $-\gamma$, namely 
\begin{align}\label{ACsigmaKp}
\sigma^{K'}_{xy}(q,N_L,\gamma,\Omega)=\begin{cases}
\sigma^K_{xy}(q,N_L,-\gamma,\Omega)& \text{if } \quad E_{\text{F}}>\Delta \\
-\sigma^K_{xy}(q,N_L=1,\gamma,\Omega)& \text{if} \quad |E_{\text{F}}|<\Delta,
\end{cases}
\end{align}
which immediately yields the closed expression for the total nonlocal Hall conductivity as 
\begin{align}\label{ACsigmaTot1}
&\sigma_{xy}(q,N_L,\gamma,\Omega)=g_{\text{s}}\bigg[\sigma^{K}_{xy}(q,N_L,\gamma,\Omega)+\sigma^{K'}_{xy}(q,N_L,\gamma,\Omega)\bigg]\nn \\
&=g_{\text{s}}\times\begin{cases}
\sigma^K_{xy}(q,N_L,\gamma,\Omega)+\sigma^K_{xy}(q,N_L,-\gamma,\Omega)& \text{if } \quad E_{\text{F}}>\Delta \\
0& \text{if} \quad |E_{\text{F}}|<\Delta,
\end{cases}
\end{align}
where $g_{\text{s}}=2$ is the spin multiplicity.

{Using the small-$q$ expansion of the $K$-valley Hall conductivity expressed in Eq. \eqref{ACsigmaK} and \eqref{ACsigmaKgfcs}, and further simplifications, the AC total nonlocal Hall conductivity for $E_{\text{F}}>\Delta$ is found to be given by
\begin{widetext}
\begin{align}\label{ACsigmaTot2}
\sigma_{xy}(q,N_L,\gamma,\Omega)& \simeq g_{\text{s}}\bigg[\sigma_{xy}(q=0,N_L,\gamma,\Omega)+\sigma^{(2a)}_{xy}(q,N_L,\gamma,\Omega)+\sigma^{(2b)}_{xy}(q,N_L,\gamma,\Omega)+\sigma^{(2c)}_{xy}(q,N_L,\gamma,\Omega)\bigg],\nn \\
\text{where}\nn \\
\sigma_{xy}(q=0,N_L,\gamma,\Omega)&=(-2\sigma_0)\frac{\Omega ^2-2N_L+1}{ \Omega ^4-2 \Omega ^2  \left(2 \gamma ^2+2 N_L-1\right)+1}\nn \\
\sigma^{(2a)}_{xy}(q,N_L,\gamma,\Omega)&=(\sigma_0q^2\ell^2)(1-N_L)\frac{\Omega ^2-2N_L+2}{ \Omega ^4-4 \Omega ^2  \left(\gamma ^2+N_L-1\right)+4}\nn \\
\sigma^{(2b)}_{xy}(q,N_L,\gamma,\Omega)&=(-\sigma_0q^2\ell^2)(N_L)\frac{\Omega ^2-2N_L}{ \Omega ^4-4 \Omega ^2  \left(\gamma ^2+N_L\right)+4}\nn \\
\sigma^{(2c)}_{xy}(q,N_L,\gamma,\Omega)&=(\sigma_0q^2\ell^2)(2N_L-1)\frac{\Omega ^2-2N_L+1}{ \Omega ^4-2 \Omega ^2  \left(2\gamma ^2+2N_L-1\right)+1}.
\end{align}
\end{widetext}
\begin{figure}[!htb]
\includegraphics[angle=0,width=0.35 \linewidth]{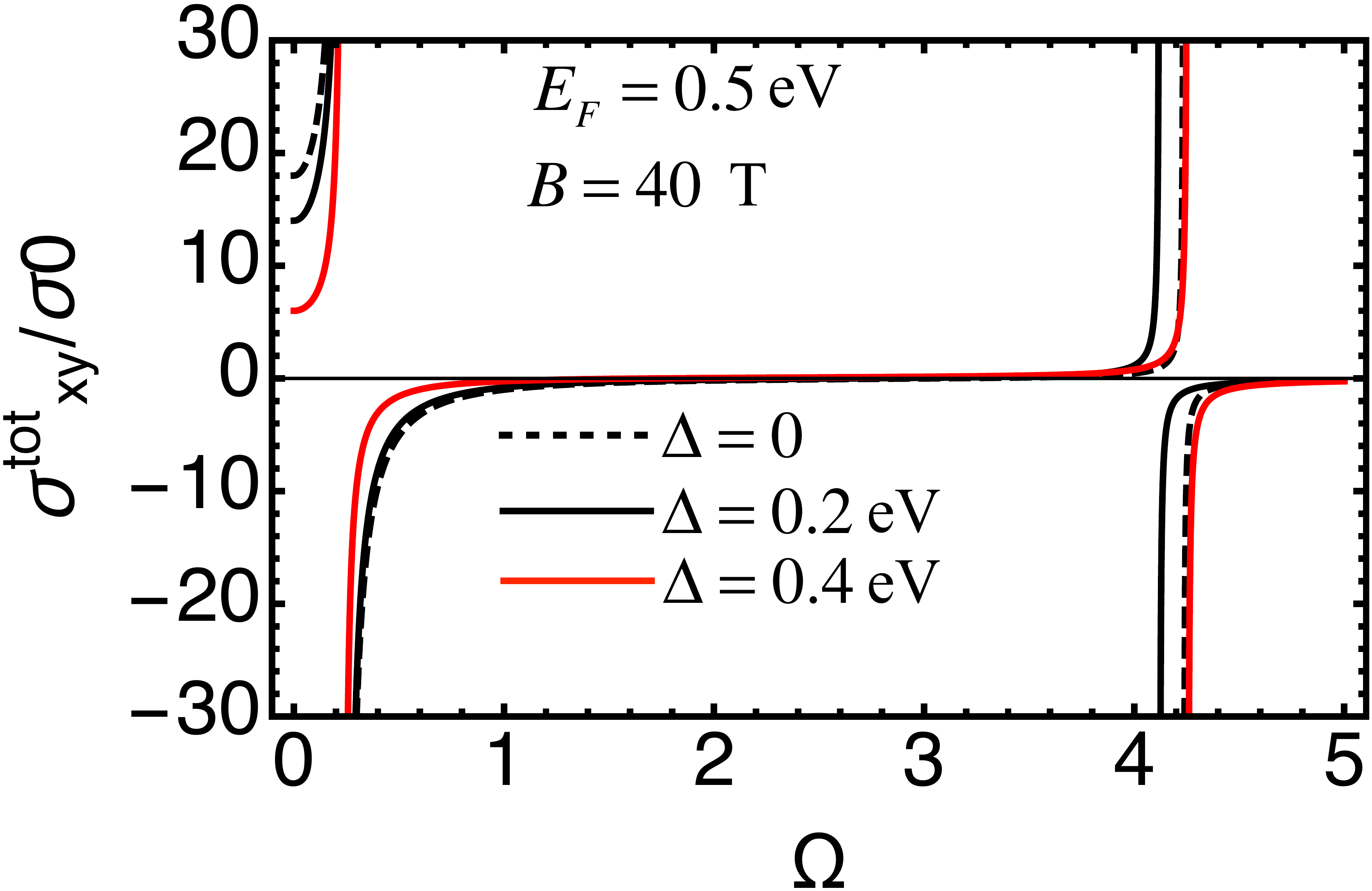}
\caption{Frequency dependence of the total nonlocal Hall conductivity for various values of the gap.} 
\label{Fig4}
\end{figure}

Eq. \eqref{ACsigmaTot2} is another central result of this paper. Fig. \ref{Fig4} shows the frequency dependence of the total Hall conductivity in Eq. \eqref{ACsigmaTot2} as a function of the frequency ratio $\Omega$ for various magnitudes of the gap. One of the striking features of this relationship is its dependence on the gap energy ($\gamma=\Delta/(\hbar\omega_0)$-dependence) which only appears in the second term in the denominators and disappears at the DC limit corresponding to $\Omega=\omega/\omega_0=0$. In other words, regardless of the magnitude of the energy gap, the DC Hall conductivity of the gapped graphene is the same as that for the gapless case including both valleys. Explicitly, evaluating Eq. \eqref{ACsigmaTot2} for $\Omega=0$ yield the DC total Hall conductivity as
\begin{widetext}
\begin{align}\label{DCsigmaTot}
\sigma_{xy}(q,N_L,\Omega=0)=2\sigma^K_{xy}(q,N_L,\Omega=0)\simeq g_{\text{s}}\frac{e^2}{h}\bigg[2N_L-1-\frac{q^2\ell^2}{4}\left(6N^2_L-6N_L+1\right)\bigg],
\end{align} 
\end{widetext}
}
Setting $q=0$ in Eq. \eqref{DCsigmaTot}, we immediately recover the well-known ``anomalous" Hall conductivity of electrons in gapless graphene $\sigma_{xy}(q=0,\Omega=0)=g_{\text{sv}}\frac{e^2}{h}(N_L-1/2)$ with the valley and spin multiplicity of $g_{\text{sv}}=4$ which yields odd-integer plateaus in the gapless graphene. It is noteworthy that Nguyen and Gromov has also studied the small-momentum expansion of the Hall conductivity of massless Dirac electrons in the gapless case (corresponding to $\gamma=0$) \cite{Nguyen} and our result of Eq. \eqref{DCsigmaTot} for the conductivity is in agreement with the Hall conductivity that can be obtained using Eq. {93} of their work presenting the Dirac polarization tensor. The agreement can easily be verified noting that their choice of units of $e=\hbar=c=1$ results in $e^2/h=1/(2\pi)$, and using Eq. \eqref{NL}, our label for the filled Landau level $N_L$ is related to their label $N$ as $N_L\equiv N+1$. In particular, the total conductivity in Eq. \eqref{DCsigmaTot} is twice as large the one that can be obtained from Eq. {93} of Nguyen and Gromov's paper \cite{Nguyen}. Apart from the spin multiplicity, this factor of two may be due to the fact that their calculation is for one valley as it can be construed from their Dirac Hamiltonian presented in their Eq. {\bf 78}.    

\section{Summary and discussion}
We have shown by explicit calculation and by theoretical argument  that the Hall viscosity and the Hall conductivity of  ``gapped graphene" in a uniform perpendicular magnetic field are both zero (at zero temperature) when the Fermi level lies in the gap $|E_{\text{F}}|<\Delta$, where $\Delta$ is a ``Semenoff mass".  In all other cases the two quantities are given by the same formulas that were previously derived for gapless graphene\cite{Sherafati2016}.    The most intriguing result of the analysis is that it is not possible, at least with our present method,  to calculate the Hall viscosity for a single valley (K or K').  The sum over Landau levels diverges with a strength proportional to the gap $\Delta$.  It is only when the contributions of K and K' are combined that the diverging terms cancel, leaving us with the familiar finite result.  This unexpected result seems to point to a fundamental limitation of the continuum model, when applied to the calculation of the valley-filtered viscosity.  In contrast to this, the results for the nonlocal conductivity remain well-behaved, even when they are calculated in a single valley.  An immediate consequence of this phenomenon is that the Hoyos-Son relation between nonlocal conductivity and viscosity breaks down in a single valley.  The physical interpretation of this singular behavior remains unclear at the time of this writing. 
\section*{Acknowledgment}
The work was supported by the grant No. DE- FG02-05ER46203 funded by the U.S. Department of Energy, Office of Science.

\onecolumngrid
\appendix
\section{Evaluation of the nonlocal Hall conductivity}
\label{AppA}

In this Appendix, we find the final expressions of the $K$-valley nonlocal Hall conductivity $\sigma^K_{xy}(\qv,\omega)$ as expressed in Eq. \eqref{Ksigmaq2} and evaluate it it for the electron-doped case for an arbitrary $N_L$ as an expansion over $\qv$. Similar expressions for the $K'$-valley are found corresponding to two separate cases for the Fermi level being first above the gap and then within the gap.  

\subsection{$E_{\text{F}}>\Delta$ corresponding to $N_L\geq 1 $}

We start with the symmetry properties of $I(n',n,\qv)$ integral, namely
\begin{align}\label{InnpProp}
I^*(n',n,\qv)&=I(n,n',-\qv)\nn \\
|I(n',n,\qv)|^2&=|I(n,n',-\qv)|^2=|I(n,n',\qv)|^2.		
\end{align}
Using Eq. \eqref{sigmaXYq}, the nonlocal Hall conductivity is then given by:
\begin{align}\label{Ksigmaq1}
\sigma^K_{xy}(\qv,\omega)&=\frac{e^2v_{\text{F}}^2}{A\hbar\omega}\sum_{\substack{{\lambda,\lambda'}\\{n,n'}\\{k_y,k'_y}}}\frac{\mathcal{F}^K(\lambda,n;\lambda',  n')}{\mathcal{E}^K(\lambda, n;\lambda', n')}\mathcal{I}^K(\lambda, n;\lambda', n', \qv),
\end{align}
where the three main functions with their symmetry with respect to the exchange of $\{\lambda,n\}\longleftrightarrow \{\lambda',n'\}$ are defined as 
\begin{align}\label{Symms}
\mathcal{F}^K(\lambda,n;\lambda', n')&=f(E^K_{\lambda,n})-f(E^K_{\lambda',n'})=-\mathcal{F}^K(\lambda', n';\lambda, n) \nn \\
\mathcal{E}^K(\lambda, n;\lambda', n')&=\omega+(E^K_{\lambda,n}-E^K_{\lambda',n'})/\hbar=\mathcal{E}^K(-\lambda', n';\lambda, n)\nn \\
\mathcal{I}^K(\lambda, n;\lambda', n', \qv)&=\lambda^2 a_{\lambda,n}^2a_{-\lambda',n'}^2\bigg\vert I(n',n-1,\qv)\bigg\vert^2-\lambda'^2 a_{-\lambda,n}^2a_{\lambda',n'}^2\bigg\vert I(n'-1,n,\qv)\bigg\vert^2=-\mathcal{I}^K(\lambda', n';\lambda, n, \qv).
\end{align}
The last relationship for function $\mathcal{I}^K(\lambda, n;\lambda', n', \qv)$ is found using the symmetry of $I(n',n,\qv)$ integrals expressed in Eq. \eqref{InnpProp}. Now, we use the properties expressed in Eq. \eqref{Symms} and consider the degeneracy of LLs to simplify the multiple summations in Eq. \eqref{Ksigmaq1} into separate summations over LLs in the conduction and valence bands. The final result will be Eq. \eqref{Ksigmaq2} in the main text. 

Now, similar to the electron-doped case of valley Hall viscosity we calculate the contributions to the nonlocal Hall conductivity in Eq. \eqref{Ksigmaq3} from three cases, namely 
\begin{enumerate}[label=(\roman*)]
\item transitions between the occupied levels in the positive sector labeled by $\{\lambda=+1, 1\leq n\leq N_L-1\}$ and the unoccupied levels labeled by $\{\lambda=+1, N_L\leq n'\leq N_{\text{C}}\}$, giving the first double sum in Eq. \eqref{Ksigmaq3}. This contribution in nonzero only for $N_L\geq2$. 
\item transitions between the occupied levels in the negative sector labeled by $\{\lambda=-1, 1\leq n \leq N_C\}$ and the unoccupied levels labeled by $\{\lambda=+1, N_L\leq n'\leq N_{\text{C}}\}$, giving the second double sum in Eq. \eqref{Ksigmaq3}, and
\item transitions between the occupied the zeroth Landau level labeled by $\{\lambda=-1, n=0\}$ and the unoccupied levels labeled by $\{\lambda=+1, N_L\leq n'\leq N_{\text{C}}\}$, giving the last single summation over $n'$. 
\end{enumerate}
 
Adding these three  contributions we obtain the total $K$-valley nonlocal Hall conductivity to be given by
\begin{align}\label{Ksigmaq3}
\sigma^K_{xy}(\qv,\omega)=\frac{e^2\omega_0^2}{h}\sum^{N_{\text{C}}}_{n'=N_L}\bigg[&\sum^{N_L-1}_{n=1}\frac{\mathcal{I}^K(+,n;+,n',\qv)}{\omega^2-\omega_0^2(\sqrt{\gamma^2+n}-\sqrt{\gamma^2+n'})^2}+\sum^{N_{\text{C}}}_{n=1}\frac{\mathcal{I}^K(-,n;+,n',\qv)}{\omega^2-\omega_0^2(\sqrt{\gamma^2+n}+\sqrt{\gamma^2+n'})^2} \nn \\
&+\frac{\mathcal{I}^K(-,0,+;n',\qv)}{\omega^2-\omega_0^2(\gamma+\sqrt{\gamma^2+n'})^2}\bigg].
\end{align}
Now, we note the cancellation that occurs in the second double sum in (ii). In particular, we can re-write this term as  
\begin{align*}
\sum^{N_{\text{C}}}_{n'=N_L}\sum^{N_{\text{C}}}_{n=1}f(n,n')&=\sum^{N_{\text{C}}}_{n'=N_L}\left(\sum^{N_L-1}_{n=1}f(n,n')+\sum^{N_{\text{C}}}_{n=N_L}f(n,n')\right)\nn \\
&=\sum^{N_{\text{C}}}_{n'=N_L}\sum^{N_L-1}_{n=1}f(n,n')+\sum^{N_{\text{C}}}_{n'=N_L}\sum^{N_{\text{C}}}_{n=N_L}f(n,n')
\end{align*}
where $f(n,n')$, the summand in the second double sum in (ii), is an anti-symmetric function with respect to the exchange of $n,n'$, according to Eq. \eqref{Symms}.
%
This allows us to simplify the Hall conductivity given in Eq. \eqref{Ksigmaq3} to the following final form: 
\begin{align}\label{Ksigmaq4}
\sigma^K_{xy}(\qv,\omega)=\frac{e^2\omega_0^2}{h}\sum^{N_{\text{C}}}_{n'=N_L}\Bigg[&\sum^{N_L-1}_{n=1}\left(\frac{\mathcal{I}^K(+,n;+,n',\qv)}{\omega^2-\omega_0^2(\sqrt{\gamma^2+n}-\sqrt{\gamma^2+n'})^2}+\frac{\mathcal{I}^K(-,n;+,n',\qv)}{\omega^2-\omega_0^2(\sqrt{\gamma^2+n}+\sqrt{\gamma^2+n'})^2}\right) \nn \\
&+\frac{\mathcal{I}^K(-,0;+,n',\qv)}{\omega^2-\omega_0^2(\gamma+\sqrt{\gamma^2+n'})^2}\Bigg],
\end{align}
where the third term is due to contribution from the ZLL being occupied in the $K$-valley for $N_L\geq 1$ and its numerator is found from Eqs. \eqref{Symms} and \eqref{Amplitude} to be $\mathcal{I}^K(-,0;+,n',\qv)=-a^2_{+,n'}\bigg\vert I(n'-1,0,\qv)\bigg\vert^2$.

According to Eq. \eqref{Jq}, the current density operator for $K'$-valley is given by $\hat{\bm j}^{K'}(\qv)=-ev_{\text{F}}\hat{\bm \sigma}^*e^{-i\qv.\hat{\rv}}$ with the pseudo-spin operator to be $\hat{\bm \sigma^*}=(-\hat{\sigma}_x,\hat{\sigma}_y)$. The nonlocal valley Hall conductivity in this case is calculated from the current-current response function similar to the one expressed in Eq. \eqref{Chiqomega} using the matrix elements of the current density operator with respect to the LLs in this valley represented in Eq. \eqref{LLs}. we finally obtain the following expression for the nonlocal Hall conductivity: 
\begin{align}\label{Kpsigmaq1}
\sigma^{K'}_{xy}(\qv,\omega)=\frac{e^2\omega_0^2}{h}\sum_{\substack{{\lambda,n;E_{\lambda,n}<E_{\text{F}}}\\{\lambda',n';E_{\lambda',n'}>E_{\text{F}}}}}\frac{\mathcal{I}^{K'}(\lambda, n;\lambda', n', \qv)}{\omega^2-\omega_{\lambda n,\lambda' n'}^2},
\end{align}   
where the function in the numerator of the summand is defined as 
\begin{align}\label{IKp}
\mathcal{I}^{K'}(\lambda, n;\lambda', n', \qv)&=\lambda^2 a_{-\lambda,n}^2a_{\lambda',n'}^2\bigg\vert I(n',n-1,\qv)\bigg\vert^2-\lambda'^2 a_{\lambda,n}^2a_{-\lambda',n'}^2\bigg\vert I(n'-1,n,\qv)\bigg\vert^2 \nn \\
&=\mathcal{I}^K(-\lambda, n;-\lambda', n', \qv) \nn \\
&=-\mathcal{I}^{K'}(\lambda', n';\lambda, n, \qv).
\end{align}
Similar to the result of the $K$-valley Hall conductivity given in Eq. \eqref{Ksigmaq4}, further simplification of Eq. \eqref{Kpsigmaq1} yields 
\begin{align}\label{Kpsigmaq2}
\sigma^{K'}_{xy}(\qv,\omega)=\frac{e^2\omega_0^2}{h}\sum^{N_{\text{C}}}_{n'=N_L}\bigg[&\sum^{N_L-1}_{n=1}\left(\frac{\mathcal{I}^{K'}(+,n;+,n',\qv)}{\omega^2-\omega_0^2(\sqrt{\gamma^2+n}-\sqrt{\gamma^2+n'})^2}+\frac{\mathcal{I}^{K'}(-,n;+,n',\qv)}{\omega^2-\omega_0^2(\sqrt{\gamma^2+n}+\sqrt{\gamma^2+n'})^2}\right) \nn \\
&+\frac{\mathcal{I}^{K'}(+,0;+,n',\qv)}{\omega^2-\omega_0^2(\gamma-\sqrt{\gamma^2+n'})^2}\bigg],
\end{align}
where the third term is due to contribution from the ZLL being empty in the $K'$-valley for $N_L\geq 1$ and its numerator is found from Eqs. \eqref{IKp} and \eqref{Amplitude} to be
$\mathcal{I}^{K'}(+,0;+,n',\qv)=\mathcal{I}^K(-,0;-,n',\qv)=-a^2_{-,n'}\bigg\vert I(n'-1,0,\qv)\bigg\vert^2$.

Eqs. \eqref{IKp} and \eqref{Amplitude} indicate that for $N_L\geq 1$, the result for $K'$-valley nonlocal Hall conductivity [Eq. \eqref{Kpsigmaq2}] can be found from that for the $K$-valley [Eq. \eqref{Ksigmaq4}] by the replacement $\gamma \to -\gamma$.   

\subsection{$|E_{\text{F}}|<\Delta$ corresponding to $N_L=1$ for $K$-valley and $N_L=0$ for the $K'$-valley}

In this case, we see that the ZLL for the $K$-valley is occupied ($N_L=1$) but it is empty for the $K'$-valley. Using Eq. \eqref{Ksigmaq4}, the first summation inside the bracket vanishes and the contributions from the ZLL and the transitions from all the LLs in the negative sector are then given by
\begin{align}\label{KsgmInGap}
\sigma^K_{xy}(\qv,\omega)=\frac{e^2\omega_0^2}{h}\sum^{N_{\text{C}}}_{n'=1}\frac{\mathcal{I}^K(-,0,+;n',\qv)}{\omega^2-\omega_0^2(\gamma+\sqrt{\gamma^2+n'})^2}.
\end{align}

Similarly, in Eq. \eqref{Kpsigmaq2}, the only nonzero contribution to the Hall conductivity of $K'$-valley is coming from the ZLL, namely
\begin{align}\label{KpsgmInGap}
\sigma^{K'}_{xy}(\qv,\omega)&=\frac{e^2\omega_0^2}{h}\sum^{N_{\text{C}}}_{n'=1}\frac{\mathcal{I}^{K'}(-,n';+,0,\qv)}{\omega^2-\omega_0^2(\gamma+\sqrt{\gamma^2+n'})^2},
\end{align}
%

According to \eqref{IKp} we can see that numerator in both Eqs. \eqref{KsgmInGap} and \eqref{KpsgmInGap} are opposite, namely  $\mathcal{I}^K(-,0,+;n',\qv)=-a^2_{+,n'}\bigg\vert I(n'-1,0,\qv)\bigg\vert^2$ and 
$\mathcal{I}^{K'}(-,n',+;0,\qv)=a^2_{+,n'}\bigg\vert I(0,n'-1,\qv)\bigg\vert^2$, or $\mathcal{I}^K(-,0,+;n',\qv)=-\mathcal{I}^{K'}(-,n',+;0,\qv)$ which is obvious from symmetry property of the integral $I(n',n,\qv)$ given in Eq. \eqref{InnpProp}. An important conclusion is that the total nonlocal Hall conductivity vanishes, namely $\sigma_{xy}(\qv,\omega)=\sigma^{K}_{xy}(\qv,\omega)+\sigma^{K'}_{xy}(\qv,\omega)=0$ for any Fermi level within the gap.

\subsection{Small-$q$ expansion of $\sigma^{K}_{xy}(q,\omega)$ for $E_{\text{F}}>\Delta$ corresponding to $N_L\geq 1 $}

{In this section, we expand the nonlocal Hall conductivity presented in Eq. \eqref{Ksigmaq4} up to $\mathcal{O}\left(q\ell\right)^2$. Let's  define $n'=n+k$ and write the double summation in the expression of the conductivity in terms of $n$ and $k$ variables. The upper and lower limits of the summation over $k$ are given by $\text{min}(k)=\text{min}(n')-\text{max}(n)=N_L-(N_L-1)=1$ and $\text{max}(k)=\text{max}(n')-\text{min}(n)=N_{\text{C}}-1$, respectively. For any $k$ value, the allowed $n$ values lie within the interval $\bigg[\text{max}\{1,N_L-k\},\text{min}\{N_L-1,N_{\text{C}}-k\}\bigg]$. These taken into account, $\sigma^K_{xy}(\qv,\omega)$ in Eq. \eqref{Ksigmaq4} is given by   
\begin{align}\label{KsigmaSmallq1}
\sigma^K_{xy}(\qv,\omega)=&\frac{e^2\omega_0^2}{h}\sum^{N_{\text{C}}-1}_{k=1}\sum^{\text{min}\{N_L-1,N_{\text{C}}-k\}}_{n=\text{max}\{1,N_L-k\}}\left(\frac{\mathcal{I}^K(+,n;+,n+k,\qv)}{\omega^2-\omega_0^2(\sqrt{\gamma^2+n}-\sqrt{\gamma^2+n+k})^2}+\frac{\mathcal{I}^K(-,n;+,n+k,\qv)}{\omega^2-\omega_0^2(\sqrt{\gamma^2+n}+\sqrt{\gamma^2+n+k})^2}\right) \nn \\
+&\frac{e^2\omega_0^2}{h}\sum^{N_{\text{C}}}_{n'=N_L}\frac{\mathcal{I}^K(-,0;+,n',\qv)}{\omega^2-\omega_0^2(\gamma+\sqrt{\gamma^2+n'})^2}.
\end{align}
Now, we note from Eqs. \eqref{Iq4K}, \eqref{Gq4K}, and \eqref{IK} that the numerators of the summands are functions of $q\ell$ argument with rapidly growing degree, namely for $n',k>1$ we have 
\begin{align}\label{IKnk}
&\mathcal{I}^K(\pm,n;+,n+k,\qv)=\frac{2^{-k-3}e^{-q^2\ell^2/2}}{\sqrt{\left(\gamma ^2+n\right)\left(\gamma ^2+k+n\right)}}\Bigg\{\left(\gamma \pm\sqrt{\gamma ^2+n}\right) \left(\pm\sqrt{\gamma ^2+k+n}\mp\gamma \right)\frac{(n-1)!}{(k+n)!}\bigg[(q\ell)^{k+1}L^{k+1}_{n-1}(q^2\ell^2/2)\bigg]^2\nn \\
&\pm4 \left(\gamma \mp\sqrt{\gamma ^2+n}\right) \left(\gamma +\sqrt{\gamma ^2+k+n}\right)\frac{n!}{(k+n-1)!}\bigg[(q\ell)^{k-1}L^{k-1}_{n}(q^2\ell^2/2)\bigg]^2\Bigg\}\nn \\
&\mathcal{I}^K(-,0;+,n',\qv)=-a^2_{+,n'}\bigg\vert I(n'-1,0,\qv)\bigg\vert^2=-\frac{2^{-n'}\left(\gamma+\sqrt{\gamma^2+n'}\right)}{\sqrt{\gamma^2+n'}(n'-1)!}(q\ell)^{2n'-2}e^{-q^2\ell^2/2}
\end{align}

Eq. \eqref{IKnk} signifies that up to $\mathcal{O}\left(q\ell\right)^2$, the only terms that survive in the small-$q$ expansion of Eq. \eqref{KsigmaSmallq1} correspond to $k=1$ and $k=2$ for the double summation and $n'=1,2$ for the last summation corresponding to the ZLL contribution. In other words, only a finite number of terms exist in the small-$q$ expansion of Eq. \eqref{KsigmaSmallq1}. 

Let's focus on the case for which $N_L\geq 3$. In this case the ZLL contribution is vanishing in the small-q expansion of Eq. \eqref{KsigmaSmallq1}. The conductivity expansion is then given by:
\begin{align}\label{KsigmaSmallq2}
\sigma^K_{xy}(\qv,\omega)\simeq&\frac{e^2\omega_0^2}{h}\sum^{2}_{k=1}\sum^{N_L-1}_{n=N_L-k}\left(\frac{\mathcal{I}^K(+,n;+,n+k,\qv)}{\omega^2-\omega_0^2(\sqrt{\gamma^2+n}-\sqrt{\gamma^2+n+k})^2}+\frac{\mathcal{I}^K(-,n;+,n+k,\qv)}{\omega^2-\omega_0^2(\sqrt{\gamma^2+n}+\sqrt{\gamma^2+n+k})^2}\right)\nn \\
&=\frac{e^2\omega_0^2}{h}\bigg[H_{N_L-1,1}(q,\gamma,\omega)+H_{N_L-2,2}(q,\gamma,\omega)+H_{N_L-1,2}(q,\gamma,\omega)\bigg],
\end{align}       
where, using Eq. \eqref{IKnk}, the functions $H_{n,1}$ and $H_{n,2}$ are given by
\begin{align}\label{Hnk}
H_{n,1}(q,\gamma,\omega)&=-\frac{\big[(2n+1) (q\ell)^2-2\big] \left(\sqrt{\gamma ^2+n+1} \left((2 n+1) \omega_0^2-\omega ^2\right)-\gamma  \left(\omega ^2+\omega_0^2\right)\right)}{4 \sqrt{\gamma ^2+n+1} \big[\omega ^4-2 \omega ^2 \omega_0^2 \left(2 \gamma ^2+2 n+1\right)+\omega_0^4\big]}\nn \\
H_{n,2}(q,\gamma,\omega)&=-\frac{(n+1) (q\ell)^2\left(\gamma  \left(\omega ^2+2 \omega_0^2\right)+\sqrt{\gamma ^2+n+2} \left(\omega ^2-2 (n+1)\omega_0^2\right)\right)}{4 \sqrt{\gamma ^2+n+2} \big[\omega ^4-4 \omega ^2 \omega_0^2 \left(\gamma ^2+n+1\right)+4 \omega_0^4\big]}.
\end{align} 
From Eqs. \eqref{Hnk} and \eqref{Ksigmaq2}, the $q^0$ term can be obtained as
\begin{align}\label{Ksigmasmallq0}
\sigma^K_{xy}(q=0,\gamma,\omega)\simeq \frac{e^2\omega_0^2}{h}H_{N_L-1,1}(q=0,\gamma,\omega)=\frac{e^2\omega_0^2}{h}\bigg[-\frac{\gamma  \left(\omega ^2+\omega_0^2\right)+\sqrt{\gamma ^2+N_L} \left((1-2 N_L)\omega_0^2+\omega ^2\right)}{2 \sqrt{\gamma ^2+N_L} \left(\omega ^4-2 \omega ^2 \omega_0^2 \left(2 \gamma ^2+2 N_L-1\right)+\omega_0^4\right)}\bigg]=\sigma_0F(N_L,\gamma,\Omega),
\end{align}
where the last equality can be established with minor algebra with $\sigma_0=\frac{e^2}{2h}$, $\Omega=\omega/\omega_0$, and the function $F(N_L,\gamma,\Omega)$ is defined in Eq. \eqref{ACsigmaKgfcs}. With more involved algebra, we can also obtain the $q^2$ term in the small-$q$ expansion of Eq. \eqref{KsigmaSmallq2} and eventually recover Eq. \eqref{ACsigmaK} in the main text. Finally, a careful evaluation of Eq. \eqref{KsigmaSmallq1} for special cases of $N_L=1,2$ shows that the final result expressed in Eq. \eqref{ACsigmaK} is valid for all $N_L\geq 1$.}

\twocolumngrid


\begin{thebibliography}{00} 
\bibitem{GrapheneRev} A. H. Castro Neto, F. Guinea, N. M. Peres, K. S. Novoselov, and A. K. Geim, Rev. Mod. Phys. {\bf 81}, 109 (2009); D. S. L. Abergel, V. Apalkov, J. Berashevich, K. Ziegler, and T. Chakraborty, Adv. Phys. {\bf 59}, 261 (2010).

\bibitem{Chamon2012} C. Chamon, C.-Y Hou, C. Mudry, S. Ryu, and L. Santos, Phys. Scr. {\bf T146}, 014013 (2012).

\bibitem{Semenoff1984} G. W. Semenoff, Phys. Rev. Lett. {\bf 53}, 2449 (1984); A. W. W. Ludwig, M. P. A. Fisher, R. Shankar, and G. Grinstein, Phys. Rev. B {\bf 50}, 7526 (1994).

\bibitem{GrapheneSiC} S. Y. Zhou, G.-H. Gweon, A. V. Fedorov, P. N. First, W. A. de Heer, D.-H. Lee, F. Guinea, A. H. Castro Neto, and A. Lanzara, Substrate-induced bandgap opening in epitaxial graphene, Nat. Mater. {\bf 6}, 770 (2007)
\bibitem{GraphenehBNEXP1} B. Hunt, J. D. Sanchez-Yamagishi, A. F. Young, M. Yankowitz, B. J. LeRoy, K. Watanabe, T. Taniguchi, P. Moon, M. Koshino, P. Jarillo-Herrero, and R. C. Ashoori, Massive Dirac Fermions and Hofstadter Butterfly in a van der Waals Heterostructure, Science {\bf 340}, 1427 (2013).

\bibitem{GraphenehBNEXP2} R. V. Gorbachev, J. C. W. Song, G. L. Yu, A. V. Kretinin, F. Withers, Y. Cao, A. Mishchenko, I. V. Grigorieva, K. S. Novoselov, L. S. Levitov, A. K. Geim, Detecting topological currents in graphene superlattices. Science {\bf 346}, 448 (2014); C. R. Woods, L. Britnell, A. Eckmann, R. S. Ma, J. C. Lu, H. M. Guo, X. Lin, G. L. Yu, Y. Cao, R. V. Gorbachev, A. V. Kretinin, J. Park, L. A. Ponomarenko, M. I. Katsnelson, Yu. N. Gornostyrev, K. Watanabe, T. Taniguchi, C. Casiraghi, H-J. Gao, A. K. Geim, and K. S. Novoselov. Commensurate–incommensurate transition in graphene on hexagonal boron nitride. Nat. Phys. {\bf 10}, 451 (2014).

\bibitem{GraphenehBNEXP3} Z.-G. Chen, Z. Shi, W. Yang, X. Lu, Y. Lai, H. Yan, F. Wang, G. Zhang, and Z. Li, Observation of an intrinsic bandgap and Landau level renormalization in graphene/boron-nitride heterostructures, Nat. Comm. {\bf 5}, 4461 (2014). 


\bibitem{ValleyHallEff} D. Xiao, W. Yao, and Q. Niu, Phys. Rev. Lett. {\bf 99}, 236809 (2007); R. V. Gorbachev, J. C. W. Song, G. L. Yu, A. V. Kretinin, F. Withers, Y. Cao, A. Mishchenko, I. V. Grigorieva, K. S. Novoselov, L. S. Levitov, and A. K. Geim, Science {\bf 346}, 448 (2014); M. Sui, G. Chen, L. Ma, W.-Y. Shan, D. Tian, K. Watanabe, T. Taniguchi, X. Jin, W. Yao, Di Xiao, and Y. Zhang, Nat. Phys. {\bf 11}, 1027 (2015); Y. Shimazaki, M. Yamamoto, I. V. Borzenets, K. Watanabe, T. Taniguchi, and S. Tarucha, Nat. Phys. {\bf 11}, 1032 (2015); Y. D. Lensky, J. C. W. Song, P. Samutpraphoot, and L. S. Levitov, Phys. Rev. Lett. {\bf 114}, 256601 (2015); T. Ando, J. Phys. Soc. Japan {\bf 84}, 114705 (2015).

\bibitem{Goerbig2011} M. O. Goerbig, Rev. Mod. Phys. {\bf 83}, 1193 (2011).

\bibitem{Koshino2010} M. Koshino and T. Ando, Phys. Rev. B {\bf 81}, 195431 (2010).

\bibitem{LLsGapGraph2015} W. -X. Wang, L.-J. Yin, J.-B. Qiao, T. Cai, S.-Y. Li, R. -F. Dou, J.-C. Nie, X. Wu, and L. He, Phys. Rev. B {\bf 92}, 165420 (2015).

\bibitem{Hydro} A. Tomadin, G. Vignale, and M. Polini, Phys. Rev. Lett. {\bf 113}, 235901 (2014); I. Torre, A. Tomadin, A. K. Geim, and M. Polini, Phys. Rev. B {\bf 92}, 165433 (2015); D. A. Bandurin, I. Torre, R. Krishna Kumar, M. Ben Shalom, A. Tomadin, A. Principi, G. H. Auton, E. Khestanova, K. S. Novoselov, and I. V. Grigorieva et al., Science {\bf 351}, 1055 (2016); J. Crossno, J. K. Shi, K. Wang, X. Liu, A. Harzheim, A. Lucas, S. Sachdev, P. Kim, T. Taniguchi, and K. Watanabe et al. {\bf 351}, 1058 (2016); L. Levitov and G. Falkovich, Nat. Phys. (London) {\bf 12}, 672 (2016); A. Principi, G. Vignale, M. Carrega, and M. Polini, Phys. Rev. B {\bf 93}, 125410 (2016).

\bibitem{Avron} J. E. Avron, R. Seiler, and P. G. Zograf, Phys. Rev. Lett. {\bf 75}, 697 (1995); J. E. Avron, J. Stat. Phys. {\bf 92}, 543 (1998).

\bibitem{Read2009} N. Read, Phys. Rev. B {\bf 79}, 045308 (2009).

\bibitem{WenZee} X. G. Wen and A. Zee, Phys. Rev. Lett. {\bf 69}, 953 (1992).

\bibitem{Biswas2016} R. R. Biswas and D. T. Son, Proc. Nat. Acad. Sci. {\bf 113}, 8636 (2016).

\bibitem{PhotonicHallVis} N. Schine, A. Ryou, A. Gromov, A. Sommer, and J. Simon, Nature (London) 534, 671 (2016); N. Schine, M. Chalupnik, T. Can, Tankut, A. Gromov, and J. Simon, Nature {\bf 565}, 173 (2019).

\bibitem{HallVisco} N. Read, Phys. Rev. B {\bf 79}, 045308 (2009); I. V. Tokatly and G. Vignale, J. Phys. Condens. Matter 21, 275603 (2009); N. Read and E. H. Rezayi, Phys. Rev. B {\bf 84}, 085316 (2011); B. Bradlyn, M. Goldstein, and N. Read, Phys. Rev. B 86, 245309 (2012); C. Hoyos, Int. J. Mod. Phys. B {\bf 28}, 1430007 (2014).

\bibitem{HoyosSon2012} C. Hoyos and D. T. Son, Phys. Rev. Lett. {\bf 108}, 066805 (2012).

\bibitem{Burmistrov2019} I. S. Burmistrov, M. Goldstein, M. Kot, V. D. Kurilovich, and P. D. Kurilovich, Phys. Rev. Lett. {\bf 123}, 026804 (2019).

\bibitem{Sherafati2016} M. Sherafati, A. Principi, and G. Vignale, Phys. Rev. B {\bf 94}, 125427 (2016).

\bibitem{TokatlyVignale2007} I. V. Tokatly and G. Vignale, 76, 161305 (2007); (E) 79, 199903 (2009).


\bibitem{Vozmediano10}  M. A. H. Vozmediano, M. I. Katsnelson, and F. Guinea, 
Phys. Rep. {\bf 496}, 109 (2010).

\bibitem{Golkar2014}S. Golkar, M. M. Roberts, and D. T. Son, J. High Energy Phys. {\bf 12}, 138 (2014). 

\bibitem{Tuegel2015}T. I. Tuegel and T. L. Hughes, Phys. Rev. B {\bf 92}, 165127 (2015).

\bibitem{Kimura2010}  T. Kimura, arXiv:1004.2688.

\bibitem{ReadRezayi2011}N. Read and E. H. Rezayi, Phys. Rev. B {\bf 84}, 085316 (2011).

\bibitem{Gusynin2007Rev} P. Gusynin, S. G. Sharapov, and J. Carbotte, Int. J. Mod. Phys. B {\bf 21}, 4611 (2007).

\bibitem{GR} I.S. Gradshteyn and I.M. Ryzhik, \textit{Table of integrals, series and products}, Eighth edition (Academic Press, Burlington, 2007), 8.97, 7.377.

\bibitem{AQHGraph} 
	K. S. Novoselov, A. K. Geim, S. V. Morozov, D. Jiang, M. I. Katsnelson, I. V. Grigorieva, S. V. Dubonos, and  A. A. Firsov, Nature {\bf 438}, 197 (2005); 
	Y. Zhang, Y.-W. Tan, H. L. Stormer, and P. Kim, {\it ibid}, 201 (2005);
	V. P. Gusynin and S. G. Sharapov,
	Phys. Rev. Lett. {\bf 95}, 146801 (2005).
	
\bibitem{Nguyen} D. X. Nguyen and A. Gromov, Phys. Rev. B. {\bf 95}, 085151 (2017).







\end{thebibliography}
\end{document}